\documentclass[12pt]{article}

\input epsf
\usepackage{amssymb,amsmath,wrapfig}
\usepackage[matrix,arrow,curve]{xy}
\usepackage{cite}
\usepackage{graphicx}

\makeatletter
\@addtoreset{equation}{section}
\makeatother

\def\rr {{\Bbb R}}
\def\cc {{\Bbb C}}
\def\pp {{\Bbb P}}
\def\zz {{\Bbb Z}}
\def\del {\partial}

\def\del {\partial}

\def\sut {{${\rm SU}(3)$}}
\def\vol {\mathrm{vol}}
\def\sla#1{\rlap{\begin{picture}(10,10)
\put(0,0){\line(1,1){10}}
\end{picture} }#1}

\def\del          {\partial}

\DeclareMathOperator{\re}{Re}
\DeclareMathOperator{\im}{Im}

\def\sqr#1#2{{\vcenter{\vbox{\hrule height.#2pt
 \hbox{\vrule width.#2pt height#1pt \kern#1pt \vrule width.#2pt}\hrule
 height.#2pt}}}}

\def\beq {\begin{equation}}
\def\bea {\begin{eqnarray}}
\def\eeq {\end{equation}}
\def\eea {\end{eqnarray}}

\def\re{\mbox{Re }}
\def\im{\mbox{Im }}

\def\del{\partial}

\def\CN {{\cal N}}

\def\CO {{\cal O}}

\def\lsim{\mathrel{\rlap{\lower4pt\hbox{\hskip1pt$\sim$}}
    \raise1pt\hbox{$<$}}}                % less than or approx. symbol
\def\gsim{\mathrel{\rlap{\lower4pt\hbox{\hskip1pt$\sim$}}
    \raise1pt\hbox{$>$}}}                % greater than or approx. symbol

\newcommand{\Tr}{{\rm Tr\,}}

\begin{document}

            \begin{titlepage}

\begin{flushright}
WIS/08/10-JULY-DPPA\\
\end{flushright}

            \begin{center}

            \vskip .3in \noindent

            {\Large \bf{Massive type IIA string theory\\ \vskip .1in cannot be strongly coupled}}

            \bigskip

            Ofer Aharony$^1$, Daniel Jafferis$^2$,
     Alessandro Tomasiello$^{3,4}$ and Alberto Zaffaroni$^3$\\

            \bigskip

            {\small $^1$ Department of Particle Physics and Astrophysics\\
            The Weizmann Institute of Science, Rehovot 76100, Israel
            \vspace{.1cm}

            $^2$ School of Natural Sciences,
        Institute for Advanced Study, Princeton, NJ 08540, USA\\
        \vspace{.1cm}

            $^3$ Dipartimento di Fisica, Universit\`a di Milano--Bicocca, I-20126 Milano, Italy\\
            and\\
            INFN, sezione di Milano--Bicocca,
            I-20126 Milano, Italy \vspace{.1cm}
            %\bigskip

        $^4$ Albert Einstein Minerva Center, Weizmann Institute of Science, Rehovot 76100, Israel}

            \vskip .5in
            {\bf Abstract }
            \vskip .1in

            \end{center}
            %\vskip .4in

            \noindent
Understanding the strong coupling limit of massive type IIA string theory is a longstanding problem. We argue that perhaps this problem does not exist; namely, there may be no strongly coupled solutions of the massive theory. We show explicitly that massive type IIA string theory can never be strongly coupled in a weakly curved region of space-time. We illustrate our general claim with two classes of massive solutions in AdS$_4\times \cc\pp^3$: one, previously known, with ${\cal N}=1$ supersymmetry, and a new one with ${\cal N}=2$ supersymmetry. Both solutions are dual to $d=3$ Chern--Simons--matter theories.
In both these massive examples, as the rank $N$ of the gauge group is increased, the dilaton initially increases in the same way as in the corresponding massless case; before it can reach the M--theory regime, however, it enters a second regime, in which the dilaton decreases even as $N$ increases.
 In the ${\cal N}=2$ case, we find supersymmetry--preserving gauge--invariant monopole operators whose mass is independent of $N$. This predicts the existence of branes which stay light even when  the dilaton decreases. We show that, on the gravity side, these states originate from D2--D0 bound states wrapping the vanishing two--cycle of a conifold singularity that develops at large $N$.

            \vfill
            \eject

            %\addtocontents{toc}
            %\tableofcontents

            \end{titlepage}

\section{Introduction and summary of results} % (fold)
\label{sec:intro}

One of the most striking aspects of string theory is its uniqueness, realized by the famous ``web of dualities'' that interconnect its various perturbative realizations. A famous thread in this web connects weakly coupled, perturbative type IIA string theory with its strong coupling
limit, M theory (which reduces at low energies to eleven--dimensional supergravity).

It has been known for a while, however, that this duality does not work when the Romans mass parameter $F_0$ \cite{romans}, which can be thought of as a space-filling Ramond-Ramond (RR) 10-form flux, is switched on. There is no candidate parameter in eleven--dimensional supergravity to match with $F_0$, unlike for all the other fluxes; nor is there any massive deformation of the eleven--dimensional theory \cite{sagnotti-tomaras,bautier-deser-henneaux-seminara,deser} \footnote{See \cite{Bergshoeff:1997ak,Bergshoeff:1998bs} for some attempts to lift massive type IIA string theory to eleven dimensions.}. And, from the type IIA point of
view, the D0-branes which give rise to the momentum modes in the eleventh dimension at strong coupling do not exist in the
massive theory (as there is a tadpole for their worldvolume gauge field).
This would then appear to be an imperfection in our understanding of string duality: it would be one string theory whose strong coupling limit is not known.

In this paper, we will argue that this strong coupling limit may not exist, and we will show this explicitly
at least at the level of weakly curved solutions. In general these are the only solutions we have any
control over, unless we have a large amount of supersymmetry; one can separately consider cases with a
large amount of supersymmetry, and none of them seem to lead to strong coupling either. (The type I' theory
of \cite{polchinski-witten} contains in some of its vacua strongly coupled regions of massive type IIA
string theory, but these regions have a varying dilaton and their size is never larger than the string scale.) Thus, we claim that
there is no reason to believe that any strongly coupled solutions exist (with the exception of solutions with small strongly coupled regions), and we conjecture that there are none. This is consistent with the fact that no suggestion for an alternative description of the massive theory at strong coupling is known.

In section \ref{sec:gen} we provide a simple argument that the string coupling $g_s$ in massive type IIA
string theory must be small, if the curvature is small. Generically, we find that $g_s \lsim l_s/R$, where
$R$ is a local radius of curvature. The argument just uses the supergravity equations of motion and flux quantization.

This result is in striking contrast with what happens in the massless case. In the ten--dimensional massless vacuum, for example, the dilaton is a free parameter, and in particular it can be made large, resulting in the M--theory phase mentioned earlier. The massive theory has no such vacua.

It is of interest to consider examples with AdS$_4$ factors, where we can take advantage of a
dual field theory interpretation via the AdS/CFT correspondence, which also provides a non-perturbative definition for         the corresponding string theory backgrounds. In particular, it is natural to consider solutions like the   ${\cal N}=6$ supersymmetric solution  AdS$_4\times \cc\pp^3$  of the massless type IIA string theory \cite{nilsson-pope,watamura-2,sorokin-tkach-volkov-11-10}.  In this solution, the dilaton is determined by the internal flux integers $k\propto \int_{\cc\pp^1} F_2$ and $N \propto \int_{\cc\pp^3} F_6$, $g_s \sim N^{1/4}/k^{5/4}$, whereas the curvature radius $R/l_s \sim N^{1/4}/k^{1/4}$. In particular, for $N\gg k^5$ one has a large dilaton with small curvature. In this limit, the solution
is better described as the AdS$_4\times S^7/Z_k$ M--theory background. The dual field theory has been identified in \cite{abjm} as the ${\cal N}=6$ superconformal Chern-Simons-matter theory with gauge group $U(N)\times U(N)$ and Chern-Simons couplings $k$ and $-k$.

%It is natural to try and compare the massless and massive results more closely. In particular,
Massive type IIA solutions are also known on AdS$_4\times \cc\pp^3$, and it is natural to compare their behavior
to the massless case. For example, some solutions with ${\cal N}=1$ supersymmetry are known explicitly \cite{t-cp3,koerber-lust-tsimpis}; they contain the ${\cal N}=6$ solution as a particular case. The field theory duals are Chern--Simons--matter theories whose levels do not sum up to zero. Even though $F_0$ is quantized as $n_0/(2\pi l_s)$, one might think that introducing the smallest quantum of it, say $n_0=1$, should have little effect on the solutions, if the other flux integers $k$ and $N$ are already very large. It would seem, then, difficult to understand how a massless solution with large dilaton can suddenly turn into a massive solution with small dilaton when $n_0$ is turned on.

As we will see in section \ref{sec:n1}, in general this ``small deformation'' intuition is flawed. When trying to express the dilaton in terms of the flux parameters, in the massive case one ends up with expressions in which $F_0$ multiplies other, large flux parameters. Hence $F_0$ can have a large effect on the behavior of the solutions even if it is the smallest allowed quantum. As it turns out, as we increase $N$, the dilaton does start growing as $g_s \sim N^{1/4}/k^{5/4}$, as in the massless case. But, before it can become large, $g_s$ enters a second phase, where it starts decreasing with $N$. Specifically, for $N$ larger than the ``critical value'' $k^3/n_0^2$, we have $g_s\sim N^{-1/6} n_0^{-5/6}$. Both behaviors are visible in figure \ref{fig:gs}.

Notice that what happens for these ${\cal N}=1$ solutions is not entirely a consequence of the general argument in section \ref{sec:gen}. One could have found, for example, that for large $N$ the radius of curvature became small in string units. In such a situation, our supergravity argument would not have been able to rule out a large dilaton; even worse, it would actually generically predict it to be large. It is interesting to ask whether there are situations where that happens. Of course, one would not trust such strongly--curved, strongly--coupled solutions, since we have no control over them; but, if they existed, they would suggest that perhaps strongly coupled solutions
do exist and need to be understood.

To look for such a different behavior, we turn to a second class of massive solutions, still on AdS$_4 \times \cc\pp^3$, but this time with ${\cal N}=2$ supersymmetry. Such gravity solutions were predicted to exist via AdS/CFT \cite{gaiotto-t}, and found as first--order perturbations in $F_0$ of the ${\cal N}=6$ solution in \cite{abjm}. The field theory duals are again Chern--Simons--matter theories whose levels do not sum up to zero. In section \ref{sec:mono} we point out that these theories have certain gauge--invariant monopole operators, whose mass (which is protected by supersymmetry) is independent of the rank $N$. This suggests the existence of wrapped branes that remain light in the large $N$ limit. This cannot happen for backgrounds which are both weakly--coupled and weakly--curved.

To see what happens at large $N$, in section \ref{sec:n2} we find these ${\cal N}=2$ gravity solutions, generalizing the construction in \cite{petrini-zaffaroni} (see also \cite{Lust:2009mb}). We reduce the equations of motion and supersymmetry equations to a system of three ODEs for three functions, which we study numerically. As in section \ref{sec:n1}, we then study the behavior of $g_s$ as a function of the flux integers. We find exactly the same phenomenon as in section \ref{sec:n1}: $g_s$ follows initially the same growth observed for the ${\cal N}=6$ solutions, and departs from that behavior before it can get large. The existence of the light states found in section \ref{sec:mono} is not a consequence of strong coupling, but is instead explained by the fact that the internal space develops a conifold singularity where branes can wrap a small cycle.  We compute numerically the mass of D2--D0 bound states wrapping the vanishing cycle, and we reproduce very accurately the mass predicted in section \ref{sec:mono} from AdS/CFT.

Hence, in both examples we examined, the curvature stays bounded almost everywhere, and the dilaton does not become strongly coupled \footnote{A correlation between the string
coupling and the curvature in massive type IIA string theory was noticed recently in \cite{Singh:2009tq}.}. Our argument in section \ref{sec:gen} does not rule out the possibility of solutions with large curvature and large dilaton, and it would be nice to find a way to rule them out. In general, such solutions would not be trustworthy, but in some situations one might understand them via chains of dualities. For example, in some cases it might be possible to T--dualize to a massless solution with small curvature, which in turn might be liftable to M--theory, along the lines of \cite{hull-romansmass}.  The behavior found in the two examples analyzed in this paper may not be universal, and we expect the AdS/CFT correspondence to be very helpful in any further progress.

One motivation for understanding the strong coupling limit of massive type IIA string theory is the Sakai--Sugimoto model \cite{sakai-sugimoto} of holographic QCD, which has $N_f$ D8--branes separating a region of space with $F_0=0$ from a region with $F_0=N_f/(2\pi l_s)$. The solution of this model is known in the IR, where it is weakly coupled and weakly curved and the D8--branes may be treated as probes; but it is not clear what happens in the UV, where, before putting in the D8--branes, the coupling became large (see \cite{burrington-kaplunovsky-sonnenschein} for an analysis of the leading order back-reaction of the D8--branes in this model). Our analysis rules out the possibility that the region of massive type IIA string theory between the D8--branes becomes strongly coupled while remaining weakly curved in the UV. It would be interesting to understand whether there is a sensible UV completion of this model, and, if so, what it looks like.

% section intro (end)

\section{A general bound on the dilaton} % (fold)
\label{sec:gen}

In this section, we will find a bound for the dilaton for type IIA solutions with non-zero 0-form flux $F_0\neq 0$, assuming that the ten--dimensional curvature is small.

The argument is simply based on the equations of motion of type IIA supergravity. Note that due to
supersymmetry, these equations are actually exact (at two-derivative order) and can be trusted even when the
coupling constant becomes large.
%As a shorthand, let us define, for a $k$--form $\alpha= \frac1{k!}\alpha_{M_1 \ldots M_k} dx^{M_1}\wedge \ldots \wedge dx^{M_k}$,
%\begin{equation}
%   (\alpha \circ \alpha)_{MN} \equiv \frac1{(k-1)!} \alpha_{m m_2 \ldots m_k} \alpha_n{}^{m_2 \ldots m_k}\ ,
%\end{equation}
%as well as the internal product
%\begin{equation}\label{eq:cdot}
%       \alpha^2 \equiv (\alpha \cdot \alpha) \equiv
%       \frac1{k!}\alpha_{m_1 \ldots m_k}\alpha_{n_1 \ldots n_k}g^{m_1 n_1}\ldots g^{m_k n_k}\ .
%\end{equation}
The Einstein equations of motion in the string frame take the form
\begin{equation}\label{eq:eom}
    e^{-2 \phi} \left(R_{MN} + 2 \nabla_M \nabla_N \phi -\frac14 H_M{}^{PQ} H_{NPQ}\right) = \sum_{k=0,2,4} T^{F_k}_{MN}\ ,
\end{equation}
where
\begin{equation}
    T^{F_k}_{MN} = \frac1{2(k-1)!} F_M{}^{M_2\ldots M_k}F_{NM_2\ldots M_k} -\frac1{4k!} F_{M_1\ldots M_k}F^{M_1\ldots M_k} g_{MN}\ .
\end{equation}
The equations (\ref{eq:eom}) are valid at every point in
spacetime, away from possible branes or orientifolds. On such
objects, we would need to include further localized terms, but
they will not be needed in what follows. In fact, all we need is a
certain linear combination: let us multiply (\ref{eq:eom}) by
$e_0{}^M e_0{}^N$, where $e$ are the inverse vielbeine; $0$ is a
frame index in the time direction. We can now use frame indices to massage $T_{00}$ on
the right hand side:
\begin{equation}
\begin{split}
    2T^{F_k}_{00}&= \frac1{(k-1)!}
    F_0{}^{A_2 \ldots A_k} F_{0A_2\ldots A_k} - \eta_{00}
    \left(
    \frac1{2(k-1)!} F^{0 A_2 \ldots A_k}F_{0 A_2 \ldots A_k} +
    \frac1{2k!}F^{A_1 \ldots A_k} F_{A_1 \ldots A_k}
    \right)\\
    &=\frac1{2(k-1)!}
    F_0{}^{A_2 \ldots A_k} F_{0A_2\ldots A_k}
    +\frac1{2k!}F^{A_1 \ldots A_k} F_{A_1 \ldots A_k}\equiv\frac12(F_{0,k-1}^2 + F_{\perp,k}^2)\ .
\end{split}
\end{equation}
We have defined the decomposition $F_k = e^0 \wedge F_{0,k-1} + F_{\perp,k}$. (In particular, $F_{\perp,0}$ is simply $F_0$.) Applying this to (\ref{eq:eom}), we get
\begin{equation}\label{eq:arg}
    e^{-2 \phi} \left[e_0{}^M e_0{}^N\left(R_{MN} + 2 \nabla_M \nabla_N \phi -\frac14 H_M{}^{PQ} H_{NPQ}\right)\right]=
    \frac14 (\sum_{k=2,4} F_{0,k-1}^2 +\sum_{k=0,2,4} F_{\perp,k}^2)\ .
\end{equation}
Again, this is satisfied at every spacetime point (away from
possible sources): there is no integral in (\ref{eq:arg}).
$R_{MN}$ needs to be small in the supergravity approximation. In
fact, all the remaining terms in the parenthesis on the left-hand side need
to be small too: they are all two--derivative NS--NS terms. If any
of them is large in string units, we cannot trust the
two--derivative action any more; hence that parenthesis needs to
be $\ll l_s^{-2}$.

On the other hand, when $F_0 \neq0$, the right-hand side of (\ref{eq:arg}) is at least of order one in string units. To see this, recall that RR fluxes are quantized, in appropriate sense. The $F_k$ are actually not closed under $d$, but under $(d-H\wedge)$. 
However, the fluxes
\begin{equation}
    \tilde F_k = \left[ e^{-B}(F_0 + F_2 + F_4 + F_6 + F_8 + F_{10}) \right]_k
\end{equation}
are closed; when integrated over a closed space-like cycle $C_a$, they satisfy the quantization law
\begin{equation}\label{eq:fq}
    \int_{C_a} \tilde F_k = n_k (2\pi l_s)^{k-1}\ ,
\end{equation}
where $n_k$ are integers.
In particular, $F_0=n_0/(2\pi l_s)$. Since the right-hand side of (\ref{eq:arg}) is a sum of positive terms, we get that it is $> 1/l_s^2$ (up to irrelevant order one factors).

Let us now put these remarks together. Since the parenthesis on the left-hand side is $\ll 1/l_s^2$, and the right-hand side is $>1/l_s^2$, we have
\begin{equation}\label{eq:gsmall}
    e^{\phi}\ll 1\ .
\end{equation}

For generic solutions, the parenthesis on the left hand side of (\ref{eq:arg}) will be of order $1/R^2$, where $R$ is a local radius of curvature. In that case, we can estimate, then,
\begin{equation}
    e^\phi \lsim \frac{l_s}{R}\ ,
\end{equation}
which of course agrees with (\ref{eq:gsmall}).

When $F_0 = 0$, the conclusion (\ref{eq:gsmall}) is not valid because all the remaining terms on the right hand side can be made small, in spite of flux quantization. For example, assume all the components of the metric are of the same order $1/R^2$ everywhere, and that $H=0$. Then, the integral of $F$ is an integer $n_k$, but the value of $F^2_k$ at a point will be of order $(n_k/R^k)^2$ (in string units). At large $R$, this can be made arbitrarily small. This is what happens in most type IIA flux compactifications with $F_0 = 0$; the dilaton can then be made large, and the limit $\phi\to\infty$ reveals a new phase of string theory, approximated by eleven--dimensional supergravity.

To summarize, we have shown that $F_0\neq0$ implies that the dilaton is small (\ref{eq:gsmall}), as long as the two--derivative action (the supergravity approximation) is valid.

% section gen (end)

\section{The ${\cal N}=1$ solutions} % (fold)
\label{sec:n1}

In this section, we will see how the general arguments of section \ref{sec:gen} are implemented in the ${\cal N}=1$ vacua of \cite{t-cp3}.

\subsection{The ${\cal N}=1$ solutions} % (fold)
\label{sub:n1sol}

We recall here briefly the main features of the ${\cal N}=1$ solutions in \cite{t-cp3} on ${\rm AdS}_4 \times \cc\pp^3$.

The metric is simply a product:
\begin{equation}\label{eq:10dN1}
    ds^2_{{\cal N}=1}= ds^2_{{\rm AdS}_4} + ds^2_{\cc\pp^3,\,{\cal N}=1}\,.
\end{equation}
Topologically, $\cc\pp^3$ is an $S^2$ fibration over $S^4$. We use this fact to write the internal metric as
\begin{equation}\label{eq:1metric}
    ds^2_{\cc\pp^3,\,{\cal N}=1} = R^2\Big(\frac18
    (d x^i + \epsilon^{ijk}A^j x^k)^2 +
    \frac1{2 \sigma} ds^2_{S^4}\Big)\,,
\end{equation}
where $x^i$ are such that $\sum_{i=1}^3 (x^i)^2=1$, $A^i$
are the components of an SU(2) connection on $S^4$
(with $p_1 =1 $), and $ds^2_{S^4}$ is the round metric on $S^4$
(with radius one).
$R$ is an overall radius, related to the AdS radius by
\begin{equation}\label{eq:Lr1}
    R_{\rm AdS}\equiv L = \frac R2 \sqrt{\frac5 {(2 \sigma + 1 )}}\ .
\end{equation}
The parameter $\sigma$ in (\ref{eq:1metric}) is in the interval $\left[2/5,2\right]$; this implies, in particular, that $L/R$ is of order 1 for these ${\cal N}=1$ solutions.
For $\sigma=2$, (\ref{eq:1metric}) is the usual Fubini--Study
metric, whose isometry group is SU(4) $\simeq$ SO(6).
For $\sigma\neq 2$, the
isometry group is simply the SO(5) that rotates the base $S^4$.

The metric (\ref{eq:10dN1}) depends on the two parameters $L$ and $\sigma$. A third parameter in the supergravity solution is the string
coupling $g_s$. Yet another parameter comes from the $B$ field. For $2/5 < \sigma < 2$, supersymmetry requires the NS-NS 3-form $H$ to be non--zero (see \cite[Eq.~(2.2)]{t-cp3}). One can solve that
constraint by writing
\begin{equation}\label{eq:BJ}
    B=-\frac{\sqrt{(2-\sigma)(\sigma-2/5)}}{\sigma+2}\, J+ \beta\
\end{equation}
where $\beta$ is a closed two--form \cite[Eq.~(4.5)]{t-cp3}. Because of gauge invariance $B \cong B + d \lambda_1$, the space of such $\beta$ is nothing but the second de Rham cohomology of the internal space, $H^2({\Bbb CP}^3)=\rr$. So we have one such parameter, which we can take to be the integral of $\beta$ over the generating two--cycle in $H_2$,
\begin{equation}\label{eq:bbeta}
    b\equiv \frac1{(2\pi l_s)^2}\int_{\cc\pp^1} \beta\ ,
\end{equation}
where we normalized $b$ so that large gauge transformations shift it by an integer.

To summarize, the ${\cal N}=1$ supergravity solutions depend on the four parameters $(L,\sigma,g_s, b)$.

% subsection n1sol (end)

\subsection{Inverting the flux quantization equations} % (fold)
\label{sub:fq1}

We now apply the flux quantization conditions (\ref{eq:fq}). It is convenient to separate the contribution from the zero--mode $\beta$:
\begin{equation}
    \tilde F_k = e^{-\beta} \tilde F_k|_{\beta=0}\ ;
\end{equation}
we then define $\int\tilde F_k|_{\beta=0}\equiv  n^b_k (2 \pi l_s)^{k-1}$, which can be computed explicitly \cite{t-cp3}. We have
\begin{equation}\label{eq:fluxq}
    \left(\begin{array}{c}\vspace{.3cm}
        \frac 1{l g_s}f_0(\sigma)\\ \vspace{.3cm}
        \frac l{g_s} f_2(\sigma)\\\vspace{.3cm}
        \frac{l^3}{g_s} f_4(\sigma)\\
        \frac{l^5}{g_s} f_6(\sigma)
    \end{array}\right)
    =
        \left(\begin{array}{c}\vspace{.3cm}
            n^b_0 \\ n^b_2\vspace{.3cm} \\ n^b_4 \vspace{.3cm}\\ n^b_6
        \end{array}\right)
       \equiv
   \left(\begin{array}{cccc}\vspace{.3cm}
       1 & 0 & 0 & 0\\\vspace{.3cm}
       b & 1 & 0 & 0\\\vspace{.3cm}
       \frac12 b^2 & b & 1 & 0 \\
       \frac16 b^3 &  \frac12 b^2 & b & 1
   \end{array}\right)
   \left(\begin{array}{c}\vspace{.3cm}
       n_0 \\\vspace{.3cm} n_2 \\\vspace{.3cm} n_4 \\ n_6
   \end{array}\right)\ ,
\end{equation}
where
\begin{equation}\label{eq:l}
    l=L/(2\pi l_s)\ ,
\end{equation}
and
\begin{equation}
    \begin{split}
        f_0(\sigma)= \frac54 \sqrt{\frac{(2-\sigma)(5 \sigma-2)}{(2 \sigma+1 )}}\ ,\qquad
        &f_4(\sigma)= -\frac{2^5 \pi^2}{3\cdot 5^2}\frac{(\sigma-1)(2 \sigma+1)^{5/2}}{\sigma^2 (\sigma+2)^2}\sqrt{(2-\sigma)(5 \sigma -2) }\ ,\\
        f_2(\sigma)= \frac{8\pi}{\sqrt{5}}\frac{(\sigma-1)}{(\sigma+2)}\sqrt{2\sigma+1}\ ,\qquad
        &f_6(\sigma)=-\frac{2^7 \pi^3}{3 \cdot 5^{7/2}}\frac{(\sigma^2 -12 \sigma -4)(2 \sigma+1)^{7/2}}{\sigma^2 (\sigma+2)^2} \ .
    \end{split}
\end{equation}
Equation (\ref{eq:fluxq}) is \cite[Eq.~(4.26)]{gaiotto-t}, which in this paper we chose to reexpress in terms of $l$ (the AdS radius in string units) rather than $r$ (the internal size in string units), to harmonize notation with section \ref{sec:n2}.

We want to invert these formulas and get expressions for the parameters $(l,g_s,\sigma, b)$ in terms of the flux integers $n_i$, as explicitly as possible. If one assumes $b=0$, this is easy \cite{t-cp3}; with $b\neq 0$, it is a bit more complicated. A good strategy is to consider combinations of the flux integers that do not change under changes of the $b$ field: in addition to $n_0$, two other combinations are
\begin{equation}
    (n_2^b)^2 - 2 n_0 n_4^b  = n_2^2 - 2 n_0 n_4 \ ,\qquad
    (n_2^b)^3 + 3 n_0^2 n_6^b -3 n_0 n_2^b n_4^b = n_2^3 + 3 n_0^2 n_6 -3 n_0 n_2 n_4\ .
\end{equation}
We then find
\begin{align}
    \label{eq:flinv1}
        &n_2^2 - 2 n_0 n_4= (f_2^2 - 2 f_0 f_4)
        \left(\frac {l} {g_s} \right)^2 =
        \frac{16}{15}\pi^2 \left(\frac {l} {g_s} \right)^2 \frac{(\sigma-1)(4 \sigma^2 -1)}{\sigma^2}\,,\\
    \label{eq:flinv2}
        & n_2^3 + 3 n_0^2 n_6 -3 n_0 n_2 n_4 = (f_2^3+ 3 f_0^2 f_6
        -3 f_0 f_2 f_4) \left(\frac {l} {g_s} \right)^3=
        \frac{8\pi^3}{5^{3/2}} \left(\frac {l} {g_s} \right)^3\frac{(-6+17 \sigma -6 \sigma^2)(2 \sigma+1)^{3/2}}{\sigma^2}      \ .
\end{align}

\begin{figure}[h]
    \centering
        \includegraphics[width=40em]{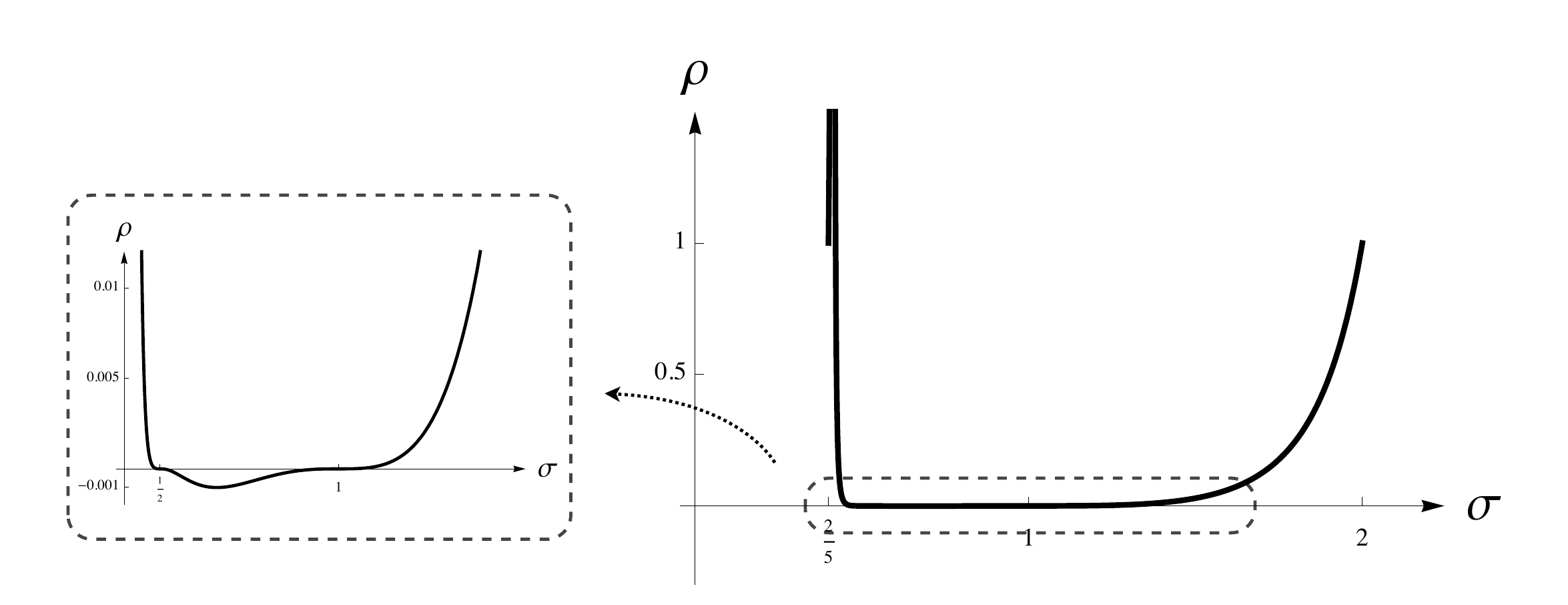}
    \caption{A plot of the function $\rho(\sigma)$ in (\ref{eq:sigma}).}
    \label{fig:rho}
\end{figure}

We see that (\ref{eq:flinv1}) and (\ref{eq:flinv2}) give two independent expressions for $l/g_s$; this implies
\begin{equation}\label{eq:sigma}
    \frac{(n_2^2 - 2 n_0 n_4)^3}{(n_2^3 + 3 n_0^2 n_6 -3 n_0 n_2 n_4 )^2}= \frac{64 (\sigma-1)^3 (2 \sigma-1)^3}{27 \sigma^2 (-6 + 17 \sigma - 6 \sigma^2)^2} \equiv \rho(\sigma) \ .
\end{equation}
This determines $\sigma$ implicitly in terms of the fluxes. The function $\rho(\sigma)$ (which we plot in figure \ref{fig:rho}) diverges at $\sigma=\frac{17-\sqrt{145}}{12}\sim .41$, and has zeros at $\sigma=\frac12$ and $\sigma=1$. These zeros have multiplicity three, and hence they are also extrema and inflection points. Moreover, it has a minimum at $\sigma\sim .65$; and it goes to 1 for both $\sigma=2$ and $\sigma=\frac25$.

We can now combine the equation for $n_0$ in (\ref{eq:fluxq}), which determines $g_s l$, with the expression for $l/g_s$ in either (\ref{eq:flinv1}) or (\ref{eq:flinv2}). We prefer using the latter, since it turns out to contain functions of $\sigma$ which are of order one on most of the parameter space:
\begin{align}\label{eq:r}
    l &= \frac {5^{3/4}}{2^{3/2}\sqrt{\pi}}
    \frac{(2-\sigma)^{1/4}(5 \sigma-2)^{1/4}\sigma^{1/3}}{(2 \sigma+1)^{1/2}(-6 + 17 \sigma - 6 \sigma^2)^{1/6}}
    \left( \frac{n_2^3}{n_0^3} + 3 \frac{n_6}{n_0}- 3 \frac{n_2 n_4}{n_0^2} \right)^{1/6}\ ,\\
    \label{eq:gs}
    g_s & = 5^{1/4}
    \sqrt{\frac\pi 2}
    \frac{(2-\sigma)^{1/4}(5 \sigma-2)^{1/4}(-6 + 17 \sigma - 6 \sigma^2)^{1/6}}
    {\sigma^{1/3} n_0^{\frac12}
    (n_2^3 + 3 n_0^2 n_6 -3 n_0 n_2 n_4)^{1/6}}\ .
\end{align}
The function in the expression for $l$ diverges at $\sigma=\frac{17-\sqrt{145}}{12}\sim .41$ and vanishes for $\sigma=\frac 25$ and 2, whereas the function in the expression for $g_s$ vanishes for $\sigma=\frac25$, $\frac{17-\sqrt{145}}{12}$ and 2.

Finally, the second row of equation (\ref{eq:fluxq}) determines $b$ in terms of $n_2$, $n_0$ and the remaining fields $l$, $g_s$ and $\sigma$. One could eliminate $l$ and $g_s$ from that expression using (\ref{eq:r}) and (\ref{eq:gs}), but we will not bother to do so.

% subsection fq1 (end)

\subsection{A phase transition} % (fold)
\label{sub:regime1}

We will start by taking for simplicity
\begin{equation}
     n_4=0 \ ,
\end{equation}
and we will call
\begin{equation}
    n_2 \equiv k \ ,\qquad n_6 \equiv N
\end{equation}
as in \cite{abjm}.

In this case, (\ref{eq:sigma}) reads
\begin{equation}
    \rho(\sigma) =\left(1+3 \frac {N n_0^2} {k^3}\right)^{-2}\ .
\end{equation}
From the graph in figure \ref{fig:rho}, we see that the behavior of the solution depends crucially on the ratio $\frac{N n_0^2}{k^3}$. If for example
\begin{equation}
    N \ll \frac{k^3}{n_0^2}\ ,
\end{equation}
we have $\rho(\sigma)\sim 1$. Looking at figure \ref{fig:rho}, we see that a possible solution is $\sigma=2$.
Around this point, $\rho$ goes linearly; so, if we write $\sigma=2- \delta \sigma$, we have $\delta \sigma \sim \frac {N n_0^2}{k^3}$. From (\ref{eq:r}) and (\ref{eq:gs}) we then have
\begin{equation}\label{eq:as1}
    l\sim \delta \sigma^{1/4}\left(\frac{k}{n_0}\right)^{1/2}= \frac{N^{1/4}}{k^{1/4}}\ ,\qquad
    g_s \sim \delta \sigma^{1/4} (k n_0)^{-1/2}=  \frac{N^{1/4}}{k^{5/4}}\ .
\end{equation}
This is the same behavior as in the ${\cal N}=6$ solution \cite{abjm}.

If, on the other hand,
\begin{equation}
    N \gg \frac {k^3}{n_0^2}\ ,
\end{equation}
we have $\rho(\sigma)\sim 0$. The possible solutions are $\sigma
\simeq 1$ or $\sigma \simeq \frac{1}{2}$. The
$\sigma$--dependent functions in the expressions for $l$ and $g_s$
in (\ref{eq:r}) and (\ref{eq:gs}) are then both of order one. We have
\begin{equation}\label{eq:as2}
    l\sim \frac{N^{1/6}}{n_0^{1/6}}\ ,\qquad g_s \sim\frac1{N^{1/6} n_0^{5/6} }\ .
\end{equation}

Notice that this behavior occurs for example in the nearly
K\"ahler solutions of \cite{behrndt-cvetic}. For those vacua, we
have $l^5/g_s = n_6$ and $1/(l g_s)= n_0$, which gives the same
behavior as in (\ref{eq:as2}). Notice also that $\sigma=1$
corresponds indeed to a nearly K\"ahler metric.

If one were to find a Chern--Simons dual to a vacuum whose only
relevant fluxes are $n_6$ and $n_0$, such as the nearly K\"ahler
solutions, it would be natural to identify $n_6$ with a rank $N$
and $n_0$ with a Chern--Simons coupling $\tilde k$ (because $F_0$
induces a Chern--Simons coupling on D2--branes). In such a dual,
$\frac{n_6}{n_0}=\frac N{\tilde k}\equiv \tilde \lambda$ would
then be the new 't Hooft coupling. We see then that $l$ and $g_s
N$ in (\ref{eq:as2}) are both functions of this $\tilde \lambda$,
as expected.

From (\ref{eq:as2}) one can calculate the finite temperature free energy to
be $\beta F \sim V_2 T^2 \frac{N^2}{\tilde\lambda^{1/3}} \sim V_2
T^2 N^{5/3} n_0^{1/3}$, which grows with a higher power of $N$
than in the massless case, for which at strong coupling
$\beta F \sim V_2 T^2 N^{3/2} k^{1/2}$.

\begin{figure}[h]
    \centering
        \includegraphics[width=25em]{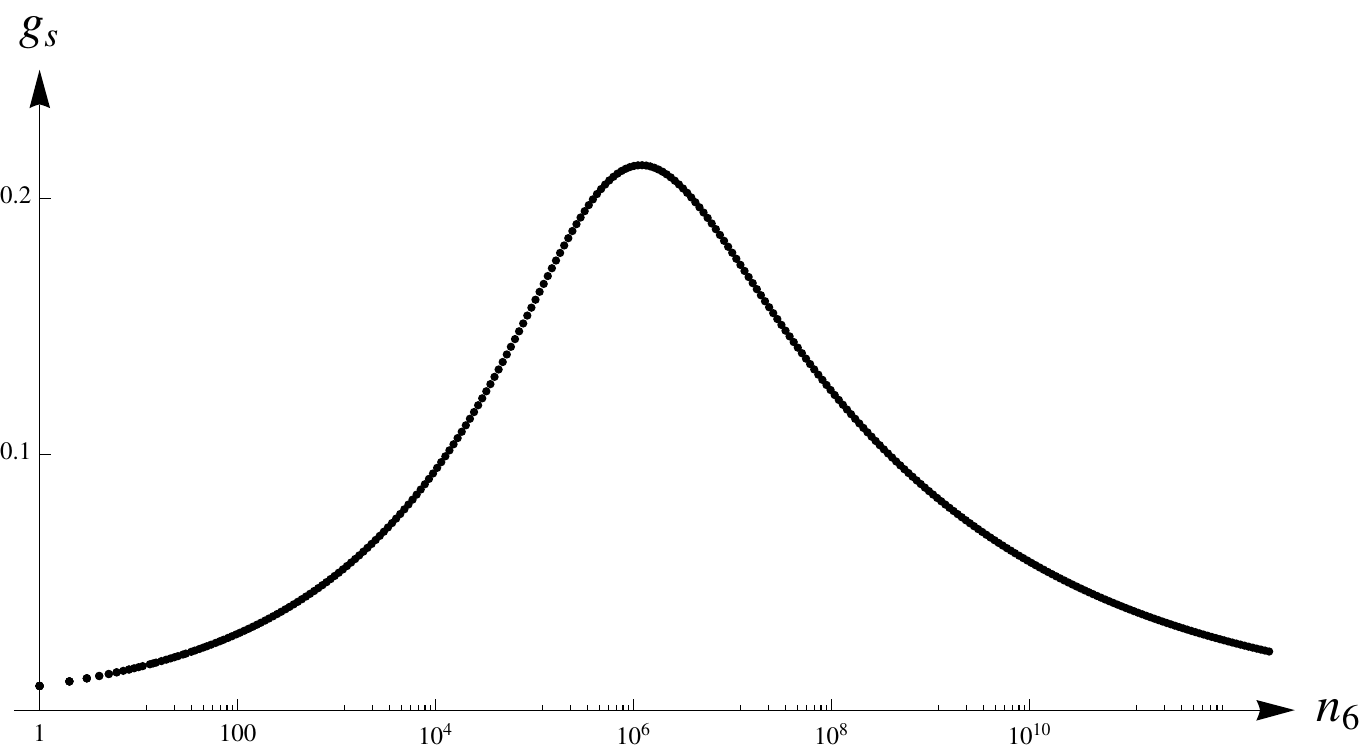}
    \caption{The behavior of $g_s$ as a function of $N=n_6$, for $n_2=k=100$, $n_4=0$ and $n_0=1$. We see both the growth in the first phase (\ref{eq:as1}), for $n_6 \ll n_2^3/n_0^2 = 10^6$, and the decay in the second phase (\ref{eq:as2}), for $n_6 \gg n_2^3/n_0^2 = 10^6$.}
    \label{fig:gs}
\end{figure}

In figure \ref{fig:gs} we show a graph of $g_s$ as a function of $N$; we see both behaviors (\ref{eq:as1}) and (\ref{eq:as2}).

% subsection regime1 (end)

Our analysis above was limited for simplicity to the case $n_4=0$, but it is easy to argue that also for
other values of $n_4$, $g_s$ cannot become large. Equation (\ref{eq:gs}) tells us that $g_s = f(\sigma) /
n_0^{1/2} m^{1/6}$, where $f(\sigma)$ is bounded from above in the relevant range of values, and
$m \equiv n_2^3 + 3 n_0^2 n_6 - 3 n_0 n_2 n_4$ is an integer. Thus, if $m \neq 0$, then $g_s$ is clearly bounded
from above in the massive theory by the maximal value of $|f(\sigma)|$. If $m=0$, then (\ref{eq:flinv2})
implies that $(-6+17\sigma-6\sigma^2)$ also vanishes, and we can then use (\ref{eq:sigma}) to rewrite
(\ref{eq:gs}) in the form $g_s = {\tilde f}(\sigma) / n_0^{1/2} {\tilde m}^{1/4}$, where ${\tilde f}(\sigma)$
is again bounded in the relevant range and ${\tilde m} \equiv n_2^2 - 2 n_0 n_4$ is another integer. Thus, if $\tilde m \neq 0$ then $g_s$ is bounded from above by
the maximal value of $|{\tilde f}(\sigma)|$, but this must be true since $m$ and ${\tilde m}$ cannot vanish
at the same time (as is clear from (\ref{eq:flinv1}) and (\ref{eq:flinv2})). Thus, for any integer fluxes with $n_0 \neq 0$,
$g_s$ is bounded from above by a number of order one.

\subsection{Probes} % (fold)
\label{sub:probes1}

We will now see that the ``phase transition'' between (\ref{eq:as1}) and (\ref{eq:as2}) has a sharp consequence on the behavior of the probe branes in the geometry. We will consider branes which are particles in ${\rm AdS}_4$ and that wrap different cycles in the internal space $\cc\pp^3$.
%We will also take
%\begin{equation}
%
%\end{equation}
%to simplify the formulas.

Not all such wrapped branes are consistent. In the ${\cal N}=6$ case, where $F_0 = 0$ and $\int_{\cc\pp^1} F_2=n_2 \neq 0$, the action for a D2-brane particle wrapping the internal $\cc\pp^1$ has a tadpole for the world--sheet gauge field ${\cal A}$, because of the coupling
\begin{equation}
    \frac{1}{2\pi l_s} \int_{\rr\times \cc\pp^1} {\cal A} \wedge F_2 = n_2 \int_{\rr} {\cal A}
\end{equation}
(the $\rr$ factor in the D2-brane worldvolume being time).  D0-branes, in contrast, have no such problem. In the field theory, they correspond \cite{abjm} to gauge--invariant operators made of monopole operators and bifundamentals.

For the solutions with both $F_0 \simeq n_0\neq 0 $ and $\int_{\cc\pp^1} F_2  \simeq n_2 \neq 0$, both D2's and D0's have a tadpole. If one considers a bound state of $n_{\rm D2}$ D2 branes and $n_{\rm D0}$ D0 branes, the tadpole for ${\cal A}$ is
\begin{equation}\label{eq:notadpole}
    (n_{\rm D2} n_2 + n_{\rm D0} n_0) \int_{\rr} {\cal A}\ .
\end{equation}
For relatively prime $n_0$ and $n_2$, the minimal choice that makes this vanish is $n_{\rm D2}= n_0$ and $n_{\rm D0}=-n_2$. These branes also correspond to a mix of monopole operators and bifundamentals;  we will discuss analogous configurations in more detail in section \ref{sub:probes2}.

 Consider now the case $n_0=1$, and $n_2 = k \gg 1$. Here we should consider a bound state of one D2 brane and $k$ D0 branes. In the context of AdS/CFT, all masses are naturally measured in units of the AdS mass scale $m_{\rm AdS} \equiv \frac1{R_{\rm AdS}}=\frac1L$; recall also from (\ref{eq:Lr1}) that $R$ is of order $L$.
The masses of a D2 and of a D0 particle would then be (setting the string scale to one)
\begin{equation}
    m_{\rm D2} L  \sim \frac{L^2/g_s}{1/L} = \frac{L^3}{g_s} \ ,\qquad
    m_{\rm D0} L  \sim \frac{1/g_s}{1/L}= \frac L {g_s}\ .
\end{equation}
 Thus, the bound states we are considering here (the particles that have no world--sheet tadpole) have a mass of order
\begin{equation}
    m_{{\rm D2}-k{\rm D0}} = \frac L{g_s}\sqrt{k^2 + L^4}\ .
\end{equation}

Which of the two terms dominates? it turns out that the answer depends on which of the two phases, (\ref{eq:as1}) or (\ref{eq:as2}), we are considering. In both phases the ratio of the two masses is a function of $\frac N {k^3}$.

A simple computation gives that, in the first phase (\ref{eq:as1}),
the D2's mass is $\sim \sqrt{N k}$, whereas the $k$ D0 branes have mass $k \times k$. The D0's dominate the mass, which then goes like $k^2 \sqrt{1 + \frac N{k^3}}$.

In the second phase, the D2's mass is $\sim N^{2/3}$, whereas the $k$ D0's mass goes like $k \times N^{1/3}$. Hence the D2 dominates. The mass then goes like $N^{2/3}\sqrt{1+ \left(\frac N{k^3}\right)^{-2/3}}$.

\medskip

Another type of branes that have no tadpole problems are D4 branes. In the field theory, these correspond \cite{abjm} to baryon operators. In AdS units, these have a mass of order $\frac{L^4/g_s}{1/L} = \frac{L^5}{g_s}$. Interestingly, this turns out to be of order $N$ in both phases (\ref{eq:as1}) and (\ref{eq:as2}), which looks reasonable for a baryon.

% subsection probes (end)

\subsection{Field theory interpretation} % (fold)
\label{sub:ft1}

The field theories dual to the vacua analyzed in this section were proposed in \cite{gaiotto-t}. Because of the low amount of supersymmetry, we do not expect to be able to make here any useful check of this duality. However, we can use our gravity results to make some predictions about those field theories, under some assumptions.

First of all, let us recall briefly the ${\cal N}=1$ field theories defined in \cite{gaiotto-t}. They are similar to the ${\cal N}=6$ theory of \cite{abjm, aharony-bergman-jafferis}, in that they also have a gauge group ${\rm U}(N_1)\times {\rm U}(N_2)$. The matter content can be organized in (complexified) ${\cal N}=1$ superfields $X^I$, $I=1,\ldots, 4$; they transform in the $(\bar N_1, N_2)$ representation of the gauge group. The biggest difference is that the Chern--Simons couplings for the two gauge groups are now unrelated: we will call them $k_1$ and $-k_2$. For $k_1 \neq k_2$, it is no longer possible to achieve ${\cal N}=6$ supersymmetry, and there are several choices as to the amount of flavor symmetry and supersymmetry that one can preserve. In this section, we consider a choice that leads to ${\cal N}=1$ supersymmetry and SO(5) flavor symmetry; in the following sections we will consider a different choice, that leads to ${\cal N}=2$ and SO(4) flavor symmetry.

This theory can be written in terms of ${\cal N}=1$ superfields; the superpotential then reads $W_{{\cal N}=1}={\rm Tr}[ c_1 X_I^\dagger X^I X_J^\dagger X^J + c_2 X_I^\dagger X^J X_J^\dagger X^I + c_3 \omega^{IK}\omega_{JL}X_I^\dagger X^J X_K^\dagger X_L]$. Notice that all the terms are manifestly invariant under Sp(2)=SO(5), as promised. When $k_1=k_2\equiv k$, the theory has ${\cal N}=6$  supersymmetry when the parameters are $c_1=-c_2=2\pi/k$, $c_3=-4\pi/k$. For $k_1\neq k_2$, this choice is no longer possible, as we already mentioned. In spite of there being only ${\cal N}=1$ supersymmetry, however, it was argued in \cite{gaiotto-t} that there still exists a choice of $c_i$ that makes the theory superconformal, as long as $k_1-k_2$ is small enough with respect to the individual $k_i$.

If we define the 't Hooft couplings
\begin{equation}
    \lambda_1 = \frac N {k_1}\ ,\qquad \lambda_2 = \frac N {-k_2}\ ,\qquad \lambda_\pm = \lambda_1 \pm
    \lambda_2 \ ,
\end{equation}
the ${\cal N}=6$ theory would correspond to $\lambda_+ = 0$. The argument in \cite{gaiotto-t} then says that there is a CFT in this space of theories if $\lambda_+\ll \lambda_-$, although at strong coupling it is difficult to quantify just how much smaller it has to be.

Let us now try to translate in terms of these field theories the ``phase transition'' we saw in section \ref{sub:regime1}. To do so, we can use the dictionary (\ref{eq:dic}) between the field theory ranks and levels on one side, and flux integers on the other. This dictionary is also valid for ${\cal N}=1$ theories \cite{gaiotto-t}. The phase transition in section \ref{sub:regime1} happens for $N\sim n_2^3/n_0^2$. Since
\begin{equation}
    \frac{k_1\mp k_2}N = \frac1 \lambda_1 \pm \frac1\lambda_2= \frac{\pm 4 \lambda_\pm}{\lambda_+^2 - \lambda_-^2}\ ,
\end{equation}
when $\lambda_+ \ll \lambda_-$ we have $n_0/N \sim \lambda_+/\lambda_-^2$, $n_2/N \sim 1/\lambda_-$. So the phase transition happens at
\begin{equation}
    \lambda_- \sim \lambda_+^2\ .
\end{equation}
In particular, the ``ABJM phase'' (\ref{eq:as1}) corresponds to $\lambda_+ \ll \sqrt{\lambda_-} $; the ``nearly--K\"ahler'' phase (\ref{eq:as2}) corresponds to $\lambda_+ \gg \sqrt{\lambda_-} $. At strong coupling, then, there is an intermediate regime where $\sqrt{\lambda_-}\ll\lambda_+ \ll \lambda_-$, where it is possible that the second phase (\ref{eq:as2}) is also described by the field theories described in \cite{gaiotto-t} and reviewed in this section.  However, given the low amount of supersymmetry, this can only be a conjecture at this point.

Rather than trying to test further this correspondence, we will now turn our attention to ${\cal N}=2$ theories, on which there is much better control.

% subsection ft1 (end)

% section n1 (end)

\section{Monopoles in ${\cal N}=2$ Chern--Simons--matter theories} % (fold)
\label{sec:mono}

In this section, we will recall some general facts about monopole operators in Chern--Simons--matter theories, and we will apply them to a particular quiver theory, similar to the ABJM theory; its gravity dual will be examined in section \ref{sec:n2}.

\begin{subsection}{Construction of monopole operators in general}
\label{sub:monogen}

Consider a $d=3$ gauge theory with gauge group $\prod_{i=1}^m
{\rm U}(N_i)$. Then there are $m$ currents, $j_i = \ast\ \Tr (F_i)$, which
are conserved by the Bianchi identity. If the theory flows to a
CFT in the IR, then these must be dimension 2 operators in the IR.
There may or may not be
%gauge invariant
operators charged under
the corresponding ${\rm U}(1)^m$ flavor symmetry; if they exist we will
call them monopole operators.

In a conformal field theory,  it is convenient to use radial quantization and consider the theory on $\mathbb{R}\times S^2$. Let us apply the state--operator correspondence to a monopole operator, with charge vector
$n_i$. This results in a state in the theory on an $S^2$, such
that $\int_{S^2} \Tr(F_i) = 2\pi n_i$.  We will denote the diagonal values of $F_i$ by $w_i^a$:
\begin{equation}\label{eq:wai}
    F_i=   {\rm diag}(w_i^1,\ldots,w_i^{N_i}){\rm vol}_{S^2}\ ;
\end{equation}
taking the trace over the gauge group ${\rm U}(N_i)$, we have that the magnetic charges are
\begin{equation}
    n_i = \sum_a w^a_i\ .
\end{equation}
We are interested in $d=3$ ${\cal N}=2$ Chern--Simons--matter theories. We can take them to be weakly interacting at short distances by adding Yang--Mills terms as regulators \cite{gaiotto-jafferis,benna-klebanov-klose,jafferis,benini-closset-cremonesi}. In such a regulated theory, there are BPS classical configurations with the gauge field as in (\ref{eq:wai}) and non--trivial values for the scalar fields. The BPS equations on $\rr^{1,2}$ include the Bogomolnyi equations $F_i = * D\sigma_i$, where $\sigma_i$ is the adjoint scalar in the vector multiplet. In $\rr \times S^2$, this equation is different because the metric needs to be rescaled, and the fields need to be transformed accordingly; the equations then read $F_i= \sigma_i \vol_{S^2}$. Notice in particular that the $\sigma_i$ are constant. There are also other BPS equations, which involve the other scalars in the theory (for explicit computations for ${\cal N}=3$ theories, see \cite[\S3.2]{benna-klebanov-klose}, and in ${\cal N}=2$ language, \cite{kim-madhu}).

After adding the regulating Yang-Mills term, $g_{YM}^2$ becomes small in the UV, so the ${\cal N}=2$ vector multiplet should be treated classically, while the chiral matter fields should not. This justifies not solving Gauss's law in describing the ``pure'' monopole operator, which then behaves as if  it were a local, non-gauge invariant chiral field \cite{benna-klebanov-klose}.

The scalar $\sigma_i$  is set by the BPS equations to be equal to the inverse Chern--Simons level times the moment maps of the matter fields,
\begin{equation}\label{Dterm}
    \sigma_i = 2\pi \frac{D_i}{k_i}\ ,
\end{equation}
and to have a spin 0 operator, one should satisfy the constraint \cite{kim-madhu}
\beq\label{comm}
\sigma_i X_{ij} = X_{ij} \sigma_j
\eeq
for any bi--fundamental field $X_{ij}$ connecting the $i$-th and $j$-th gauge group \cite{jafferis-t,martelli-sparks-3dquivers,hanany-zaffaroni}. In addition, the fields $X_{ij}$ need to satisfy the F-term equations of the ${\cal N}=2$ theory. We see from (\ref{Dterm}) that some matter fields, which should be neutral under the background ${\rm U}(1)$, are necessarily non vanishing  in the background; hence, the possibility of satisfying all equations
%,  using only fields
%which are neutral under the magnetic flux ${\rm U}(1)$,
%(since we want to
%construct a chiral primary)
gives a nontrivial constraint on the
possible BPS monopole operators.

In general, monopole operators $T$ creating such configurations at
a point will not be gauge invariant. However, they will behave
exactly like local fields. Hence, they can be combined with other
local operators ${\cal O}$, to write gauge--invariant expressions
of the schematic form
\begin{equation}\label{eq:TO}
    \Tr(T\CO)\ ,
\end{equation} where the indices are contracted as appropriate for
the representations in which the operators transform.

%There are, however, constraints on the possible gauge invariant expressions we can obtain.
Let us determine how the monopole transforms under the gauge group.  This is easy to find for the Abelian
factors of the theory. There is an obvious contribution to the electric charges of a monopole operator, from the Chern--Simons term $\sum \frac{k_i}{4\pi} \int \Tr (A_i \wedge F_i)$. This says that a monopole with magnetic charges $n_i$ will behave like a particle with charges $n_i k_i$ under the $i^{\rm th}$ electric ${\rm U}(1)$ Abelian factor.

This result  gives a constraint on the possible gauge invariant operators $\Tr( T\CO)$ we can obtain.  If all matter is in bifundamental and adjoint representations, no gauge invariant operators can be formed from monopoles that are charged under the overall U(1), since no matter field transforms under it.
Since we just computed the electric charge under the $i^{\rm th}$ U(1) to be $k_i n_i$, the charge under the overall U(1) is $\sum
k_i n_i$. Thus, if we are to form any gauge invariant
operators of the form (\ref{eq:TO}), we need to require
\begin{equation}\label{eq:skn}
    \sum_{i=1}^m k_i n_i = 0\ .
\end{equation}
This result will be useful in the theories with gauge group ${\rm U}(N_1)\times {\rm U}(N_2)$, which we will discuss in section \ref{sub:monoABJM}.

Let us now ask how the monopole will transform under the full non-Abelian group. One method to determine this is the following. One fixes a particular configuration on the sphere (breaking the gauge symmetry), one computes the charge under the whole Cartan subalgebra of the gauge group, and then one integrates over the gauge orbits. Thus monopole operators can be labelled by ${\rm U}(1)$ subgroups of the gauge groups. For the ${\rm U}(N_i)$ factor of the gauge group, the charges under this Cartan subalgebra are the $w^a_i$, with $a=1,..., N_i$, that we saw in (\ref{eq:wai}).

Therefore, a monopole associated with magnetic flux $w^a_i$ is in the representation with weight vector \cite[Sec.~4.2]{kapustin-wilsonthooft}
\begin{equation}\label{eq:weight}
    (k_i w_i^1, \ldots, k_i w_i^{N_i})
\end{equation}
of the $i^{\rm th}$ gauge group. (Our notation here is that the weight vector denotes the number of boxes in each row of a Young diagram of the representation; thus $(k,0,\ldots,0)$ is, for example, the completely symmetric representation.)

Quantization in a background of the type discussed in this section can result in anomalous contributions to the charges and energy of the state. For the non--chiral theories we we will consider in section \ref{sub:monoABJM}, there is no such a correction to the gauge charges\footnote{If the matter content is chiral, as for the theories in \cite{petrini-zaffaroni}, there will be an additional one--loop correction to the gauge charges. One way to understand this effect is that the state on the sphere has a constant value for the scalar in the vector multiplet, which gives a mass to any matter fields charged under that ${\rm U}(1)$ subgroup in which the magnetic flux lives. Integrating them out at one--loop can shift the effective Chern--Simons level in that background if the theory has chiral matter.}.
However, as we will discuss in section \ref{sub:monoABJM}, the dimension of the monopole operator is given by the energy of
the state on the sphere, which often includes a non--zero Casimir
energy.

\end{subsection}

\begin{subsection}{Dimensions and charges of the monopoles}
\label{sub:monoABJM}

We will apply in this section the results of section \ref{sub:monogen} to the Chern--Simons--matter theories with ${\cal N}=2$ and ${\cal N}=3$ given in \cite{gaiotto-t}. In particular, we will compute the dimensions of particular monopoles, which will be useful later, when comparing to the gravity solutions of section \ref{sec:n2}.

Let us first recall some details about the field theories of interest. They are similar to the ${\cal N}=6$ theory of \cite{abjm, aharony-bergman-jafferis}, in that they also have a gauge group ${\rm U}(N_1)\times {\rm U}(N_2)$, and ${\cal N}=2$ ``chiral'' superfields $A_i, B_i$, $i=1,2$; the $A_i$ transform in the $(\bar N_1, N_2)$ representation, whereas the $B_i$ transform in the $(N_1, \bar N_2)$. Just as in section \ref{sub:ft1}, the crucial difference between the ${\cal N}=6$ theory and the ${\cal N}=2$ theory is that the Chern--Simons couplings for the two gauge groups are now unrelated; we again call them $k_1$ and $-k_2$. The theories we want to consider in this section are defined by the superpotential
\begin{equation}\label{sup}
    W = {\rm Tr} (c_1 (A_i B_i)^2 + c_2 (B_i A_i)^2 ) \ .
\end{equation}
For generic $c_i$, the theory has ${\cal N}=2$ supersymmetry and ${\rm SU}(2)$ flavor symmetry.
For $c_i=\frac1{k_i}$, supersymmetry turns out to be enhanced to ${\cal N}=3$, while the flavor symmetry is still ${\rm SU}(2)$. For $c_1=-c_2=c$, the supersymmetry stays ${\cal N}=2$, but $W$ can be rewritten as
\begin{equation}\label{eq:n2}
    W= c{\rm Tr} (\epsilon^{ij} \epsilon^{kl}A_i B_k A_j B_l)\ ,
\end{equation}
which shows that the flavor symmetry is enhanced to ${\rm SU}(2)\times {\rm SU}(2)$.  This  ${\cal N}=2$ theory is dual to the gravity solution discussed in the next section.

We will now apply to these theories the discussion of section
\ref{sub:monogen} about monopole operators. The following
computation is a straightforward generalization of that done in
\cite{gaiotto-jafferis,benna-klebanov-klose} for the ${\cal N}=3$
theory. The results for the ${\cal N}=3$ and $\CN=2$ theories
appear to be identical, since the flavor symmetry guarantees that
the matter fields have the same dimensions as in the more
supersymmetric theory.

As we saw in section \ref{sub:monogen}, there are non--trivial conditions on the scalars for the monopole to be BPS.  For the theories we are considering, the conditions read
\bea\label{constr}
\begin{split}
    A \, A^\dagger - B^\dagger B =& \frac{k_1}{2\pi} \sigma_1\ , \\
    B \, B^\dagger - A^\dagger A =& -\frac{k_2}{2\pi} \sigma_2\ ,\\
    \sigma_1 A_i =& A_i \sigma_2\ ,\\
      \sigma_2 B_i =& B_i \sigma_1 \  ,
\end{split}
\eea
together with the F-term constraints coming from (\ref{sup}). In view of (\ref{eq:wai}) and $F_i=\sigma_i {\rm vol}_2 $, such equations relate the magnetic fluxes $w_1^a$ with the $w_2^a$.  The simplest  monopoles  we can consider are defined by magnetic charges $w_1^a$ which are all either $0$ or $1$: namely, $w_1=(1,\ldots,1, 0,\ldots)$ (with $n_1$ 1's) and $w_2=(1,\ldots,1,0,\ldots)$ (with $n_2$ 1's). The non-zero elements of the fields
$A_i$ and $B_i$ are $n_1\times n_2$ and $n_2\times n_1$ rectangular matrices, respectively, which are required
to satisfy  the first two lines in (\ref{constr}) and the F-term constraints.  The problem of finding appropriate vacuum expectation values for the matter fields
%neutral under the ${\rm U}(1)$ subgroup of the gauge group
%such that the moment maps determine VEV's for the scalars $\sigma_i$ in the vector multiplets required for supersymmetry
is equivalent to finding the BPS moduli space of  the generalized ${\rm U}(n_1) \times {\rm U}(n_2)$ Klebanov--Witten theory with superpotential (\ref{sup}) and Fayet-Iliopoulos (FI) parameters turned on. Many of these moduli spaces are non--empty.

Using now (\ref{eq:skn}), we see that this monopole operator can be coupled to elementary fields  in a gauge invariant way   only if
\begin{equation}\label{cons}
    n_1 k_1 = n_2 k_2\ .
\end{equation}

The matter content is non--chiral, so there are no anomalous
contributions to the gauge charges of the monopole. There is, however, a  one--loop correction to the
dimension of the operator, which  is given by \cite{benna-klebanov-klose,jafferis,benini-closset-cremonesi}
\begin{equation}
\begin{split}
    \Delta &=-\frac{1}{2} \sum_{\rm fermions} |q| R \\& = -\frac12 \left [ 2 \times 1 \left ( n_1(N_1-n_1) + n_2(N_2-n_2)\right )
    - 4\times \frac12    \left (n_1(N_2-n_2) + n_2(N_1-n_1) \right ) \right ]\\
    &= (n_1-n_2)^2 -
    (N_1-N_2)(n_1-n_2)\ ,
\end{split}
\end{equation}
where $R$ is the R-charge of the fermion  and $q$ the   charge under the U(1) subgroup specified by the  vectors $w_1=(1,.._{ n_1}..,1, 0,\ldots)$ and $w_2=(1,.._{n_2}..,1,0,\ldots)$. The various contributions arise as follows.
The four bi-fundamental fermions have R-charge $-1/2$ \footnote{Since the superpotential (\ref{sup}) must have R-charge $+2$ and there is a discrete symmetry between $A_i$ and $B_i$ we have that $R(A_i)=R(B_i)=1/2$; the
R-charge of the fermionic partners is $R(A)-1$ by supersymmetry.}. Each bi-fundamental fermion is a matrix with $N_1N_2$ entries;  the $n_1(N_2-n_2)+n_2(N_1-n_1)$ off-diagonal entries have charge $\pm 1$ under the magnetic U(1) subgroup, while the remaining entries are neutral. The two adjoint gauginos have R-charge $+1$.
They are square matrices with $N_i^2$ entries; the $2 n_i(N_i-n_i)$ off-diagonal entries of the $i$-th fermion  have charge $\pm 1$, while the other are neutral. We used the fact that in this theory, both for $\CN=3$ and $\CN=2$, the R--charges in the UV and IR are identical.

In the ${\cal N}=6$ theory, $k_1=k_2$ and it follows from (\ref{cons}) that $n_1=n_2$. The simplest monopole has just
 $w_1=w_2= (1,0,\ldots , 0)$.  We need to turn on  $A_i$ and $B_i$ fields that solve the $U(1)$ Klebanov-Witten
 theory with a FI term.  According to (\ref{eq:weight}), such a monopole transforms in the $k$-fold symmetric representation
 of $U(N_1)$ and in the conjugate  $k$-fold symmetric representation of $U(N_2)$. The monopole can combine  with  $k$ fields $A_i$ to form a gauge-invariant operator (we can analogously form a gauge-invariant operator with the conjugate
 monopole and $k$ fields $B_i$).

In the ${\cal N}=3$ and ${\cal N}=2$ theories, we cannot have $n_1=n_2$, but we can now take
 $n_1=k_2$ and $n_2=k_1$ and rectangular matrices $A_i$ and $B_i$ that solve (\ref{constr}). In general, the matrices $A_i B_j$ will not be diagonalizable. Recalling equation (\ref{eq:weight}), the monopole operator is in a representation of the gauge group with weight vectors $(k_1,.._{ k_2}.., k_1, 0,...)$ and $(k_2,.._{ k_1}.., k_2, 0, ...)$. A gauge invariant combination must include $k_1 k_2$ matter bifundamentals  (if $k_1$ and $k_2$ are not relatively prime, some operator with smaller dimension could exist).  The total dimension of the gauge-invariant operator, dressed with $k_1 k_2$ elementary fields, is then
\begin{equation}\label{eq:Danielmonopoles}
    \Delta= \frac{k_1 k_2 }{2} + (k_2 -k_1)^2 -(k_2 -k_1) (N_1-N_2) \ .
\end{equation}

Note that we have determined the vacuum expectation values of the matter fields needed to ``support'' the flux to form a BPS state on the sphere using the classical moduli space. This is justified since the Higgs branch does not receive quantum corrections. More precisely, the ring of chiral operators is the ring of algebraic functions on the moduli space. There is a natural map \cite{jafferis-t,martelli-sparks-3dquivers,hanany-zaffaroni} from the moduli space of Chern--Simons--matter theories to the moduli space of the four--dimensional Yang--Mills theory with the same field content. That moduli space cannot receive quantum corrections (aside from wavefunction renormalization which fixes the coefficients of the superpotential), and only the $S^1$ bundle over that space, associated to the dual gauge fields, is quantum corrected. This precisely corresponds to 1-loop corrections to the charges and dimensions of monopole operators, which are, however, constructed in the UV weakly coupled Yang--Mills--Chern--Simons--matter theory.

Let us summarize the results of this section.
The monopole operators that create $k_2$ units of flux for the first gauge group and $k_1$ for the second have $k_1 k_2$ bifundamental indices, and hence we can contract them with $k_1 k_2$ bifundamental fields to construct a gauge--invariant operator. Such operators have dimension given by (\ref{eq:Danielmonopoles}). In particular, they stay light when $N_1=N_2 \equiv N \to \infty$. Since in general monopole operators correspond to D--branes, this seems to indicate a limit where D--branes become light, which usually signals some sort of breakdown of the perturbative description. We will see in section \ref{sub:probes2} precisely how this happens.

\end{subsection}

% section mono (end)

\section{The $\mathcal{N}=2$ solution} % (fold)
\label{sec:n2}

We now turn to writing and studying the ${\cal N}=2$ solution predicted to exist in \cite{gaiotto-t}, and found in \cite{gaiotto-t2} at first order in $F_0$. This solution will be the gravity dual of the field theory defined by the superpotential (\ref{eq:n2}), and it will serve as another illustration of the general result of section \ref{sec:gen}.

We will start in section \ref{sub:n2sol} by reducing the equations
of motion and the supersymmetry conditions to a system of three
equations for three functions of one variable. This procedure
closely parallels \cite{petrini-zaffaroni}, where an analogous
solution for the gravity dual of the Chern-Simons theory based on
the $\mathbb{C}^3/\mathbb{Z}_3$ quiver was found. In section
\ref{sub:fq2}, we will impose flux quantization, and derive
expressions for the supergravity parameters in terms of the flux
integers; in section \ref{sub:regime2} we find, just like in
section \ref{sub:regime1}, a ``phase transition'' that prevents
the dilaton from growing arbitrarily large. Finally, in section
\ref{sub:probes2} we find light D--brane states dual to the
monopole operators discussed  in section \ref{sub:monoABJM}.

\subsection{The ${\cal N}=2$ solutions} % (fold)
\label{sub:n2sol}

The ten dimensional metric we will consider is a warped product of $AdS_4$  with a compact six-dimensional internal metric with the topology of $\mathbb{CP}^3$:
\beq
\label{metric10def}
{\rm d} s_{10}^2 = e^{2 A} {\rm d}s^2_{AdS_4}  + {\rm d}s^2_6 \ .
\eeq

As discussed in \cite{cvetic-lu-pope-cpn,gaiotto-t2}, there is a foliation of $\mathbb{CP}^3$ in copies of $T^{1,1}$, which is
in turn a $S^1$ fibration over $S^2\times S^2$. The usual Fubini--Study metric can be written as
\beq
\label{metric6CP3}
{\rm d} s^2_6 =  \frac{\cos^2 (t)}{4}  {\rm d}s^2_{S^2_1} + \frac{\sin^2(t)}{4} {\rm d}s^2_{S^2_2} +
{\rm d} t^2 + \frac{1}{16} \sin^2(2t) ({\rm d} a + A_2-A_1)^2  \, ,
\eeq
where $A_i, \, i=1,2$,  are one-form connections, with curvatures
\beq \label{eq:dAJ}
d A_i = J_i\ ,
\eeq
where $J_i$ are the volume forms of the two spheres $S^2_i$.
The coordinate $t$ parametrizes the interval $[0,\pi/2]$; all the functions in our solution (including $A$ in (\ref{metric10def})) will depend on this coordinate alone. At one end of the interval $[0,\pi/2]$, one $S^2$ shrinks; at the other end, the other $S^2$ shrinks. To make this metric regular, we take the periodicity of $a$ to be  $4\pi$.

The Fubini--Study metric is appropriate for the ${\cal N}=6$ solution, which has $F_0=0$. Once we switch $F_0$ on, as we saw in section \ref{sub:monoABJM}, AdS/CFT predicts the existence of an ${\cal N}=2$ solution with isometry group ${\rm SU}(2)\times {\rm SU}(2)\times {\rm U}(1)$ (the first two factors being the flavor symmetry which is manifest in (\ref{eq:n2}), and the third being the R--symmetry). The internal metric for such a deformed solution is then given by\footnote{One could have reparameterized the coordinate $t$ so as to set one of the functions in (\ref{metric6def}) to a constant, for example $\epsilon$, as in (\ref{metric6CP3}). We have chosen, however, to fix this reparameterization freedom in another way: by choosing the pure spinors (\ref{susyI}) to be as similar as possible to those for the ${\cal N}=6$ solution, see in particular (\ref{eq:jot}).}
\beq
\label{metric6def}
{\rm d} s^2_6 =  \frac{e^{2
B_1(t)}}{4}  {\rm d}s^2_{S^2_1} + \frac{e^{2 B_2(t)}}{4} {\rm
d}s^2_{S^2_2} + \frac{ 1}{8} \, \epsilon^2(t) {\rm d} t^2 +
\frac{1}{64} \Gamma^2(t) ({\rm d} a + A_2-A_1)^2   \, .
\eeq
Were the functions $e^{2 B_i}$ non-vanishing, we would have a metric on the total space of an $S^2$ bundle over $S^2\times S^2$. To maintain the topology of $\mathbb{CP}^3$, we require that $e^{2 B_2}$ vanishes at $t=0$ and $e^{2 B_1}$ vanishes at $t=\pi/2$. With an abuse of language, we will refer to $t=0$ as the North pole and $t=\pi/2$ as the South pole, although there is no real $S^2$ fiber. To have a regular metric, $\epsilon(t)$ and $\Gamma(t)$ must behave appropriately at the poles.

It is convenient to define the combinations
\beq w_i = 4 e^{2 B_i-2 A} \eeq
which control the relative sizes of the two $S^2$'s. As discussed in Appendix \ref{apppure}, the supersymmetry equations reduce to three coupled first order ordinary differential equations for $w_1$, $w_2$ and a third function $\psi$ which enters in the spinors:
\begin{align}\label{eq:sys}
    \psi' &= \frac{\sin (4 \psi)}{\sin (4 t)}
    \frac{C_{t,\psi}(w_1+w_2) + 2 \cos^2(2t)w_1 w_2 }
    {C_{t,\psi}(w_1+w_2)\cos^2 (2\psi)+2 w_1 w_2}\ ,\nonumber \\
    w_1'&=\frac{4 w_1}{\sin(4t)}
    \frac{C_{t,\psi}(w_1 w_2 - 2\,w_2 - 2\sin^2(2 \psi)w_1) }
    {C_{t,\psi}(w_1+w_2)\cos^2 (2\psi)+2 w_1 w_2}\ ,\\
    w_2'&=\frac{4 w_2}{\sin(4t)}
    \frac{C_{t,\psi}(w_1 w_2 - 2\,w_1 - 2\sin^2(2 \psi)w_2) }
    {C_{t,\psi}(w_1+w_2)\cos^2 (2\psi)+2 w_1 w_2}\ ,\nonumber
\end{align}
where
\begin{equation} \label{eq:Ctpsi}
    C_{t, \psi} \equiv \cos^2(2t)\cos^2(2 \psi)-1.
\end{equation}

All other functions in the metric and the dilaton are algebraically determined in terms of $w_1,w_2,\psi$:
\begin{align}
        \label{eq:eps}
        \epsilon &= \sqrt{2} e^{A} (\cot(\psi) -\tan(\psi) ) \frac{\csc^2 (2t) \, \sin (2\psi) - \cos (2\psi) \, \cot (2t)\,  \psi^\prime}{2 \sqrt{1+ \cot^2 (2t) \, \sin^2 (2\psi) }}\\
        \label{eq:Gamma}
        \Gamma &= 4 e^{A}  \frac{\sin (2 t) +\cos (2 t) \cot (2 t) \sin^2 (2\psi) }{2 \sqrt{1+ \cot^2 (2t) \, \sin^2 (2\psi) }}\\
        \label{eq:4A}
        e^{4 A} &=  -\frac{4 c}{F_0} \csc (4 t) \, \sec (2 \psi) \, \tan (2 \psi) \\
        \label{eq:3Ap}
        e^{3 A -\Phi} & = c \sec (2\psi) \sqrt{1+ \cot^2 (2t) \, \sin^2 (2 \psi) } \ .
\end{align}
Here, $c$ is an integration constant, that so far is arbitrary.
The fluxes are determined as well, and have the general form
\begin{equation}\label{eq:fluxes}
    \begin{split}
        F_2 &= k_2(t) e^{2 B_1} J_1+g_2(t) e^{2 B_2} J_2 + \tilde k_2(t) \frac{i}{2} z\wedge \bar z\ ,\\
        F_4 &= k_4(t) e^{2 B_1+2 B_2} J_1\wedge J_2+\tilde k_4(t) e^{2 B_1} \frac{i}{2} z\wedge \bar z  \wedge J_1 +\tilde g_4(t) e^{2 B_2} \frac{i}{2} z\wedge \bar z  \wedge J_2\ ,\\
        F_6 &= k_6(t) \frac{e^{2B_1+2B_2}}{16}\frac{i}{2} z\wedge \bar z \wedge J_1 \wedge J_2\ ,
    \end{split}
\end{equation}
where $\frac{i}{2} z\wedge \bar z= \frac{\epsilon \Gamma}{16\sqrt{2}} dt\wedge (da+A_2-A_1)$. The full expressions for the coefficients $k_i$, $\tilde k_i$, $g_i$, $\tilde g_i$ can be found in (\ref{eq:coeffs}). The fluxes satisfy the Bianchi identities, which require that
\beq  \label{eq:tildeF}
\tilde F \equiv  e^{-B} (F_0+F_2+F_4+F_6)
\eeq
is closed. This dictates in particular that $F_0$ is constant.

\begin{figure}[h]
    \centering
        \includegraphics[width=20em]{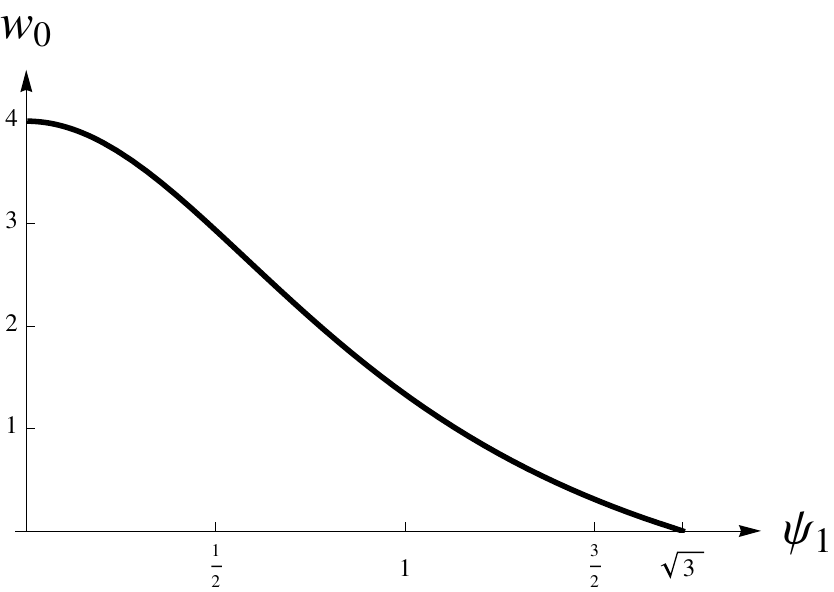}
    \caption{A plot of $w_0$ as a function of $\psi_1$. It vanishes linearly around the point $\psi_1=\sqrt{3}$.}
    \label{fig:w0}
\end{figure}

We can now study the regularity  of the differential equation near its special points, $t=0$ and $t=\pi/2$, by finding  a power series solution of the equations. The general solution will depend on three arbitrary constants. However, we are after solutions with particular topology, where $w_2$ vanishes at $t=0$ (the ``North pole") and $w_1$ vanishes at $t=\pi/2$ (the ``South pole").  Near $t=0$, we obtain
\begin{equation}
    \begin{split}
        \psi &= \psi_1 t - \frac23 (4 \psi_1 +5 \psi_1^3) t^3 + O(t^5),\\
        w_1 &= w_0 +(4+ 4 \psi_1^2 -2 w_0 +2 w_0 \psi_1^2) t^2 + O(t^4),\\
        w_2 &= (4+ 4 \psi_1^2) t^2+ O(t^4)\ .
    \end{split}
\end{equation}
In our solution, $w_0$ and $\psi_1$ are not independent: imposing that $w_1$ vanishes at $t = \pi/2$ determines $w_0$ in terms of $\psi_1$. The power series expansion in $\tilde t \equiv \pi/2-t$ near $t=\pi/2$ is identical, with the role of $w_1$ and $w_2$ exchanged:
\begin{equation}
    \begin{split}
        \psi &= \tilde \psi_1 \tilde t - \frac23 (4 \tilde \psi_1 +5 \tilde \psi_1^3) {\tilde t}^3 + O({\tilde t}^5),\\
        w_1 &= (4+ 4 \tilde \psi_1^2) {\tilde t}^2+ O({\tilde t}^4),\\
        w_2 &= \tilde w_0 +(4+ 4 \tilde \psi_1^2 -2 \tilde w_0 +2 \tilde w_0 \tilde \psi_1^2) {\tilde t}^2 + O({\tilde t}^4)\ .
    \end{split}
\end{equation}
The constants $\tilde w_0$ and $\tilde\psi_1$ should also be determined by $\psi_1$; we can then think of $\psi_1$ as the only parameter in the internal metric. To find a solution with the required topology, we note that the equations (\ref{eq:sys}) are symmetric under the operation $t\to \frac\pi2-t, \psi\to -\psi$, $w_1 \leftrightarrow w_2$, and we look for solutions which are left invariant by this symmetry. This determines $\tilde \psi_1 = \psi_1$ and $\tilde w_0 = w_0$, and it allows us to restrict the study of the equations to the ``north hemisphere'' $t\in [0,\pi/4]$. The only thing left to impose is that the solution is differentiable at $t=\pi/4$. This is what determines $w_0$ as a function of $\psi_1$, which we plot in figure \ref{fig:w0}.  $w_0(\psi_1)$ is monotonicaly decreasing; our numerical analysis shows that it vanishes at a point very well approximated by $\psi_1=\sqrt{3}$.

The perturbative expansion of the solutions near the ``poles'' $t=0$ and $t=\pi/2$ allows to check the regularity of the six-dimensional metric. In fact, the only special points in the metric are the poles,  where a copy of $S^2$ degenerates. Using the previous expansion, it is straightforward to check that, at both poles, the shrinking $S^2$ combines with $(t,a)$
to give a piece of the metric proportional to
\begin{equation}\label{eq:local}
 dt^2 + \frac14  t^2 \left ( ds^2_{S^i} + (da \pm  A_i)^2\right )\ .
\end{equation}
Thanks to the fact that the periodicity of $a$ is $4\pi$, this is the flat metric of $\mathbb{R}^4$. For all $\psi_1\in [0,\sqrt{3})$
the metric is then regular. For $\psi_1=\sqrt{3}$, both spheres degenerate at each pole and the metric develops two conifold singularities. $\psi_1=\sqrt{3}$ is thus the natural limiting point in our family of solutions.

We can examine now the number of parameters in the solution. As discussed above, the differential equations provide
just one parameter, $\psi_1$, the value of the derivative of $\psi$ at the North pole $t=0$.  It is convenient to define two more parameters by
\begin{equation}
    g_s \equiv e^\phi|_{t=0} \ ,\qquad 2 L= e^A|_{t=0}\ .
\end{equation}
 Both $\phi$ and $A$ vary over the internal manifold, but numerical study reveals that they only do so by order one functions. So $g_s$ and $L$ can be thought of as the order of magnitude of the dilaton and AdS radius in our solutions \footnote{The normalization has been chosen so that in the metric, at $t=0$, $L^2$ multiplies an Anti-de Sitter space of unit radius, and the relation between the mass of a particle at $t=0$ and the conformal dimension of the dual operator is
$m L = \Delta(\Delta -4)$. This normalization is related to the fact that, in our conventions, $ds^2_{AdS_4}$ has cosmological constant $\Lambda = -3 |\mu|^2$ and, as discussed in appendix \ref{apppure},  we chose $\mu=2$. }. We can now reexpress the integration constant $c$ by evaluating (\ref{eq:3Ap}) at $t=0$:
\begin{equation}\label{eq:c}
  c= \frac{ 8 L^3}{g_s} \frac1{\sqrt{1+\psi_1^2}}\ .
\end{equation}
The $F_0$ flux is then determined by evaluating (\ref{eq:4A}) at $t=0$:
\begin{equation}\label{eq:n0f0}
F_0 = -  \frac{1}{L g_s} \frac{\psi_1}{\sqrt{1+\psi_1^2}}\ .
\end{equation}

Finally, a fourth parameter comes from the $B$ field. As in section \ref{sec:n1}, there is a zero--mode ambiguity coming from the presence of a non--trivial cohomology in our internal manifold. To see this, let us call $B_0$ a choice of $B$--field such that $H=dB_0$ solves the equations of motion. For example, we can choose $B_0$ such that
\begin{equation}\label{eq:f2t0}
    \tilde F_2 = F_2 - B_0 F_0 = 0 \ .
\end{equation}
$H=d B_0$ is guaranteed to solve the equations of motion, since equation (\ref{eq:f2t0}) implies that $dF_2 = H F_0$, which we have already solved. However, this will also be true for any $B$ of the form
\begin{equation}\label{eq:Bb}
    B= B_0 + \beta \ ,
\end{equation}
for any $\beta$ which is closed. We can apply to this $\beta$ the same considerations as in section \ref{sub:n1sol}: because of gauge invariance $B \cong B + d \lambda_1$, the space of such $\beta$ is nothing but the second de Rham cohomology of the internal space, $H^2({\Bbb CP}^3)=\rr$, so we have one such parameter. And, just as in (\ref{eq:bbeta}), we define the integral of $\beta$ over the generating two--cycle in $H_2$: $b\equiv\frac1{(2 \pi l_s)^2} \int_{\cc\pp^1} \beta$. The fact that we use the same notation as in section \ref{sec:n1} should not generate confusion, as the contexts are different.

Summarizing, our  solutions are parameterized by the four numbers $(L,\psi_1,g_s,b)$. The situation is very similar to the ${\cal N}=1$ solutions we studied in section \ref{sec:n1}, with $\sigma$ replaced by $\psi_1$.

% subsection n2sol (end)

\subsection{Inverting the flux quantization equations} % (fold)
\label{sub:fq2}

This section will follow closely the corresponding treatment for the ${\cal N}=1$ solutions in section \ref{sub:fq1}. The equations are formally very similar:
\begin{equation}\label{eq:fluxq2}
    \left(\begin{array}{c}\vspace{.3cm}
        \frac1{l g_s} f_0(\psi_1)\\ \vspace{.3cm}
        0\\\vspace{.3cm}
        \frac{l^3}{g_s} f_4(\psi_1)\\
        \frac{l^5}{g_s} f_6(\psi_1)
    \end{array}\right)
    =
        \left(\begin{array}{c}\vspace{.3cm}
            n^b_0 \\ n^b_2\vspace{.3cm} \\ n^b_4 \vspace{.3cm}\\ n^b_6
        \end{array}\right)
       \equiv
   \left(\begin{array}{cccc}\vspace{.3cm}
       1 & 0 & 0 & 0\\\vspace{.3cm}
       b & 1 & 0 & 0\\\vspace{.3cm}
       \frac12 b^2 & b & 1 & 0 \\
       \frac16 b^3 &  \frac12 b^2 & b & 1
   \end{array}\right)
   \left(\begin{array}{c}\vspace{.3cm}
       n_0 \\\vspace{.3cm} n_2 \\\vspace{.3cm} n_4 \\ n_6
   \end{array}\right)\ ,
\end{equation}
where $l=L/(2\pi l_s)$, as in (\ref{eq:l}). The vector on the left hand side is given by the integrals $\frac{1}{(2\pi l_s)^{k-1}}\int_{C_k}\tilde F_k$, where $C_k$ is the single $k$--cycle in $\cc\pp^3$ ($k=0,2,4,6$), and $\tilde F_k$ is defined using the particular $B_0$ in (\ref{eq:f2t0}); this also explains why the second entry of the vector is zero (this is simply our choice for the definition of $b$). We could have made such a choice for the ${\cal N}=1$ solution as well; we did not do so because for \sut \,  structure solutions there is a different and particularly natural choice of $B$--field.

\begin{figure}[h]
\begin{minipage}[t]{\linewidth}
~~~\begin{minipage}[t]{0.5\linewidth}
\vspace{0pt}
\centering
\includegraphics[scale=.8]{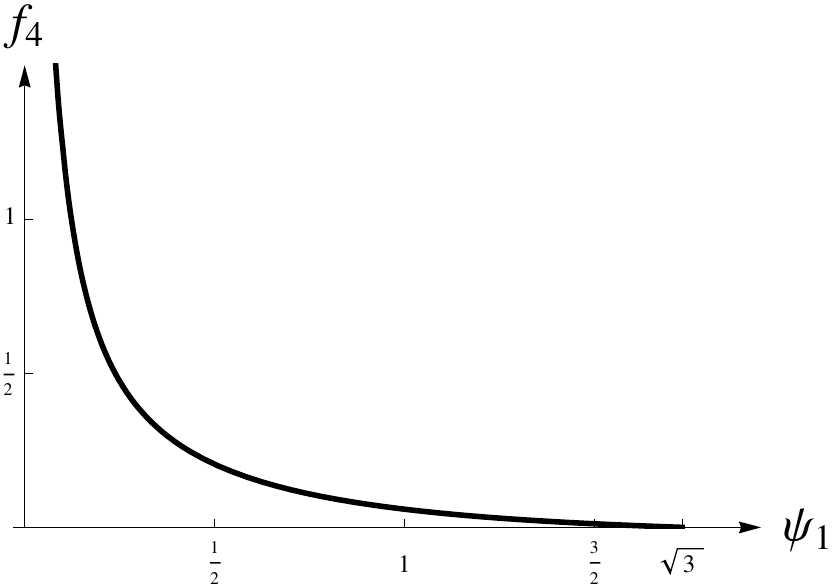}
\end{minipage}%
\hspace{.5cm}
\begin{minipage}[t]{0.5\linewidth}
\vspace{0pt}
\centering
\includegraphics[scale=.8]{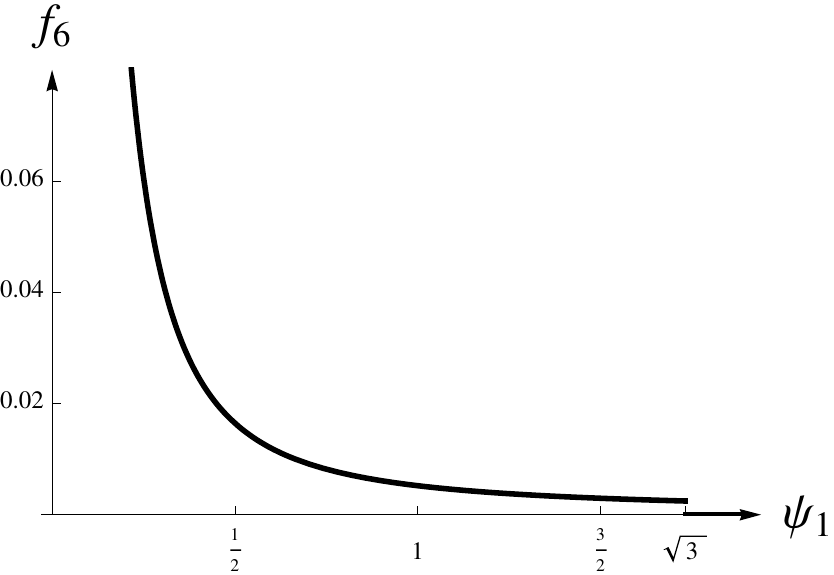}
\end{minipage}\\[1em]
\begin{minipage}[t]{0.5\linewidth}
\vspace{0pt}
\centering
\end{minipage}%
\vspace{1em}
\caption {Plots of $f_4(\psi_1)$ and $f_6(\psi_1)$. Their asymptotic behavior near $0$ and $\sqrt{3}$ is given in (\ref{eq:asf0}), (\ref{eq:asfs3}). Notice in particular that $f_6(\sqrt{3})$ is small but non--zero.}
\label{fig:fk}
\end{minipage}
\end{figure}

Using equation (\ref{eq:n0f0}) we can write
\begin{equation}\label{eq:f0}
f_0(\psi_1)= -\frac{\psi_1}{\sqrt{1+\psi_1^2}}\,.
\end{equation}
We know the other functions $f_k(\psi_1)$ only numerically. We obtain $2 f_4(\psi_1)$ by integrating $\tilde F_4$ over  the diagonal  $S^2$  times the ``fiber"  $(t,a)$, which is a representative of twice the fundamental four-cycle.
The plots of $f_k(\psi_1)$ are given in figure \ref{fig:fk}. Our numerical analysis indicates the following asymptotics at the two extrema $\psi_1=0$, $\psi_1=\sqrt{3}$:
\begin{align}\label{eq:asf0}
    f_4 &\sim \psi_1^{-1} \ ,&f_6 &\sim \psi_1^{-2}
    &{\rm for}\ \psi_1\to 0 \ ;\\
    \label{eq:asfs3}
    f_4 &\sim (\sqrt{3} - \psi_1) \ , &f_6 &\to {\rm const}
    &{\rm for}\  \psi_1 \to \sqrt{3}\ .
\end{align}

We can now proceed in the same fashion as in the ${\cal N}=1$ case to determine $\psi_1$ from the flux parameters. Namely, we write the combination
\begin{equation}\label{eq:rhoN2}
\frac{(n_2^2 - 2 n_0 n_4)^3}{(n_2^3 + 3 n_0^2 n_6 -3 n_0 n_2 n_4 )^2}= -\frac{8 f_4(\psi_1)^3} {9 f_6(\psi_1)^2 f_0(\psi_1)} \equiv \rho(\psi_1)\ ,
\end{equation}
which allows us to determine $\psi_1$ in terms of fluxes.  $l$ and $g_s$ are then given by
\bea\label{parameters}
l^6 &=& \frac{f_0(\psi_1)}{3 f_6(\psi_1)} \left (\frac{n_2}{n_0}\right )^3 \left ( 1+ 3 \frac{n_0^2 n_6}{n_2^3} - 3 \frac{ n_0 n_4}{n_2^2} \right )\ , \\
g_s^6 &=&  3 f_0^5(\psi_1) f_6(\psi_1) \left (n_2 n_0 \right )^{-3} \left ( 1+ 3 \frac{n_0^2 n_6}{n_2^3} - 3 \frac{ n_0 n_4}{n_2^2} \right )^{-1}\ .
\label{parameterstwo}
\eea

\begin{figure}[h]
    \centering
        \includegraphics[width=20em]{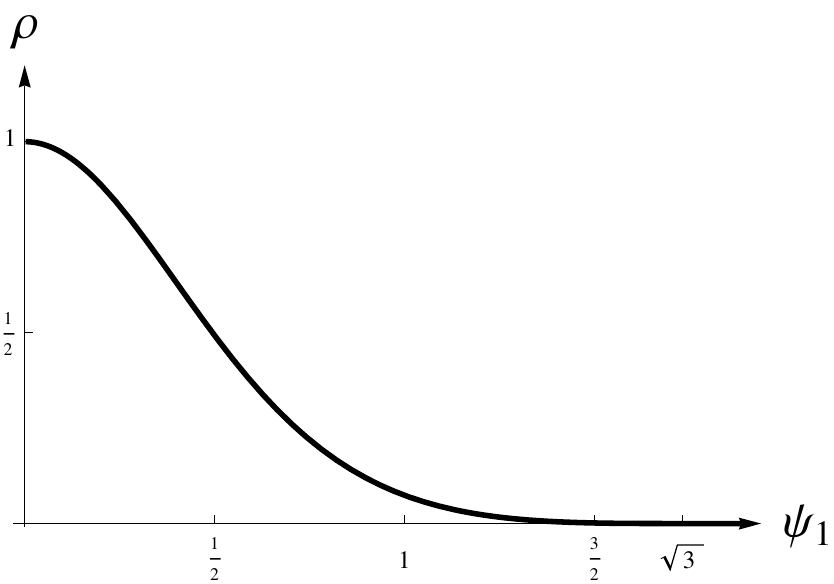}
    \caption{A plot of the function $\rho(\psi_1)$ in (\ref{eq:rhoN2}).}
    \label{fig:rhoN2}
\end{figure}

A crucial role is played by the function $\rho(\psi_1)$ which we plot in Figure  \ref{fig:rhoN2}. It decreases  monotonically from $1$ at $\psi_1=0$ to zero at $\psi_1=\sqrt{3}$. Its asymptotic behavior at $\psi_1=0$ and $\psi_1=\sqrt{3}$ is:
\begin{equation}\label{eq:asrho}
    \rho\sim 1 - {\tilde c} \,\psi_1^2 \ \ {\rm for}\ \psi_1 \to 0 \ ,\qquad
    \rho\sim (\sqrt{3}-\psi_1)^3 \ \ {\rm for}\ \psi_1 \to \sqrt{3}\ ,
\end{equation}
for some constant $\tilde c$. This is in agreement with (\ref{eq:asf0}), (\ref{eq:asfs3}). The fact that $\rho$ vanishes at the same point, $\psi_1 = \sqrt{3}$, where the solution develops a singularity, is strongly supported by our numerical analysis, and will be crucial in reproducing the field theory results.

% subsection fq2 (end)

\subsection{A phase transition} % (fold)
\label{sub:regime2}

As in the ${\cal N}=1$ case, consider for simplicity the case $n_4=0$ and call, as usual, $n_6=N$ and $n_2=k$.  Equation (\ref{eq:rhoN2})
becomes
\begin{equation}\label{eq:rhon40}
\rho(\psi_1)=\left (1+ 3 \frac{n_0^2 N}{k^3}\right )^{-2}\,.
\end{equation}

As in the ${\cal N}=1$ case, there are two interesting regimes. For $N\ll k^3/n_0^2$, $\psi_1\rightarrow 0$ and we are
near the undeformed solution. From (\ref{eq:asrho}), we see that $1-\rho(\psi_1)\sim \psi_1^2$; hence
we can identify in this regime
\begin{equation}
\psi_1\sim \left (\frac{n_0^2 N}{k^3}\right )^{1/2} \ .
\end{equation}
 Moreover, we see from (\ref{eq:f0}) that $f_0(\psi_1)\sim \psi_1$; using also (\ref{eq:asf0}), we easily compute from (\ref{parameters})
\begin{equation}
l \sim \frac{N^{1/4}}{k^{1/4}}\, , \qquad\qquad g_s \sim  \frac{N^{1/4}}{k^{5/4}}\ ,
\end{equation}
which is indeed the behavior of the ${\cal N}=6$ solution \cite{abjm}.

For $N\gg k^3/n_0^2$, the function $\rho(\psi_1)$ should approach zero, and this happens for $\psi_1\rightarrow \sqrt{3}$. From (\ref{eq:asrho}) and (\ref{eq:rhon40}), we see that
\begin{equation}
\delta\psi_1 \equiv \left ( \sqrt{3} - \psi_1\right ) \sim \left ( \frac{k^3}{N n_0^2}\right )^{2/3}\ .
\end{equation}
From (\ref{eq:asfs3}) we then conclude
\begin{equation}\label{eq:secphase2}
l\sim \frac{N^{1/6}}{|n_0|^{1/6}}\ , \qquad  g_s\sim   \frac{1}{N^{1/6}|n_0|^{5/6}}\ ,
\end{equation}
which is the same behavior as in (\ref{eq:as2}). Again, as in the ${\cal N}=1$ case, we can also argue generally
that $g_s$ remains bounded for any integer values of the fluxes, with $n_0 \neq 0$.

At first sight, this seems puzzling. At the end of section \ref{sub:monoABJM}, we noticed that this gravity solution is expected to develop light D--branes in the limit $N_1=N_2=N \to \infty$. As argued in \cite{aharony-bergman-jafferis}, $N_1=N_2$ precisely when $n_4=0$ (see also (\ref{eq:dic})). But there do not seem to be any light D--branes in a limit where $g_s$ is small and the internal manifold is large, since a D--brane mass scales as $L^k/g_s$, with $k\geq 0$.

As we remarked after equation (\ref{eq:local}), however, in the limit $\psi\to \sqrt{3}$ (which
is relevant for large $N$) the internal manifold develops two conifold--like singularities, since the two-cycle is now shrinking to zero at the ``poles". As we will now see,  the new light states are obtained from D-branes wrapping the vanishing cycle for that singularity.

% subsection regime2 (end)

\subsection{Probes} % (fold)
\label{sub:probes2}

We now want to compare this gravity solution with the field theory we saw in section \ref{sub:monoABJM}; specifically, the one defined by the superpotential in (\ref{eq:n2}), which has the right symmetries to be the dual of the gravity solution we found in section \ref{sub:n2sol}.

We can first of all try to predict what sort of bulk field corresponds to the monopole operators discussed in section \ref{sub:monoABJM}. Let us recall how the duality works in the ABJM case, when $F_0=0$. Consider first a monopole operator that creates one unit of field strength for both gauge groups at a particular point. This operator has $k$ indices under both gauge groups, and we can make it gauge-invariant by contracting it with $k$ bifundamentals. The resulting bound state corresponds to a D0 brane in the gravity dual; notice that such a brane has no tadpole on its worldsheet for the worldsheet vector potential ${\cal A}$, as we already saw in section \ref{sub:probes1}.
Another monopole operator that can be considered is the one that creates one unit of flux for, say, the second gauge group. In this case, we cannot make this operator gauge--invariant: it will have $k$ ``dangling'' indices. This corresponds to a D2 brane wrapping an internal two--cycle. As we also already saw in section \ref{sub:probes1}, such a brane has a tadpole on its worldsheet, coming from the term $\int A_1 {\cal F}= \int F_2 {\cal A}$; so one needs to have $k$ strings ending on it, and these $k$ strings correspond to the $k$ indices of the monopole operator.

When we switch on $F_0$, even a D0 brane will have a tadpole on its worldsheet, coming from the coupling $\int F_0 {\cal A}$. On the field theory side, this corresponds to the fact that the monopole operator that creates one unit of field strength for both gauge groups has now $k_1$ fundamental indices for the first gauge group and $k_2$ antifundamental ones for the second. This cannot be made gauge-invariant; we are always left with at least $|k_1-k_2|$ ``dangling'' indices. This fact was used in \cite{gaiotto-t} to establish that the Romans mass integer is the sum of the Chern--Simons couplings, so that, in the present language, $n_0= k_1 - k_2$; see also \cite{fujita-li-ryu-takayanagi,bergman-lifschytz}. In \cite{aharony-bergman-jafferis} it was similarly shown that $n_4$ is the difference between the two gauge group ranks $N_2-N_1$.\footnote{The relative sign between the expressions for $n_4$ and $n_0$ had not been determined so far. We made here a choice consistent with our final result in formula (\ref{mass}).} Putting this together, we obtain a dictionary between the flux integers and the ranks and levels of the field theory:
\begin{equation}\label{eq:dic}
    n_0 = k_1 - k_2 \ ,\qquad n_2 = k_2 \ ,\qquad
    n_4 = N_2 - N_1 \ ,\qquad n_6 = N_2 \ .
\end{equation}

In section \ref{sub:monoABJM}, we considered monopole operators which create $k_2$ units of field strength for the first gauge group, and $k_1$ units of field strength for the second. We noticed that these have $k_1 k_2$ bifundamental indices, and thus can be made gauge--invariant. Following the identifications of D2 branes and D0 branes above, if we assume for example that $k_1 > k_2$, we can say that these new gauge--invariant monopoles correspond to a bound state $k_2$ D0 branes and $k_1-k_2$ D2 branes. We have already noticed in section \ref{sub:probes1} that such a bound state can cancel the tadpole on the worldsheet, because it makes the prefactor in (\ref{eq:notadpole}) vanish.

Let us make this expectation more precise. Consider a D2 brane wrapped on a two--cycle ${\cal B}_2$ in the ${\cal N}=2$ solution. As we will see in appendix \ref{sub:bpsours}, supersymmetry requires that the D2 brane
lives at the North pole $t=0$ or at the South pole $t=\pi/2$, and that it wraps the $S^2$ that does not shrink there. We also need to cancel the tadpole for the world-volume field ${\cal A}$  which arises from the Wess-Zumino coupling,
\begin{equation}
{\cal A}\wedge \left ( F_2 + F_0({\cal F} -B)\right ) \ .
\end{equation}
We can split $B$ into a fiducial choice plus a zero mode, as in (\ref{eq:Bb}). The tadpole cancellation requires
\begin{equation}
{\cal F}- B = {\cal F}- \beta - B_0=   - F_2/F_0.
\end{equation}
Since $B_0$ was chosen to satisfy (\ref{eq:f2t0}), we need to turn on a world--volume flux
\begin{equation}
     {\cal F} = \beta \ .
\end{equation}
There is a possible obstruction to doing this, coming from the quantization of the world--volume flux, that says that $\frac{1}{(2\pi l_s)^2} \int_{S^2} {\cal F} \in \zz$. The value of $b=\frac{1}{(2\pi l_s)^2} \int_{S^2} \beta$ from (\ref{eq:fluxq2}) is given by
$b= -n_2/n_0$; hence in general it is rational and not an integer. So we see that a single D2 brane is generally not consistent. We can get around this, however, by considering $n_0$ D2--branes. In that case, the equation we want to satisfy actually reads
\begin{equation}\label{eq:Fbeta}
    {\cal F} = \beta \, 1_{n_0}\ .
\end{equation}
The integral of the trace of the left hand side is the first Chern class on the world--volume, which is the induced D0-brane charge $n_0$. The integral of the trace of the right hand side now gives $b n_0= -n_2$. We conclude that we can cancel the tadpole by considering a bound state of $n_0$ D2 branes and $n_2$ D0 branes, just as in section \ref{sub:probes1}.

Naively, one might think that the mass of a D2--D0 bound state should be at least as heavy as a D0-brane, which in units of AdS mass is $m_{D0}L\sim L/g_s$. Since this is heavy in the limit (\ref{eq:secphase2}), one might think that such a bound state can never reproduce the light mass predicted in section \ref{sub:monoABJM}.

Fortunately, such pessimism proves to be unfounded. The mass of the state is given by
\begin{equation}
m_{D2} L = n_0 L \frac{1}{ (2\pi)^2 g_s l_s^{3} }\int_{{\cal B}_2}\sqrt{ \det \left ( g+ {\cal F} -B \right ) }= n_0 L  \frac{1}{ (2\pi)^2 g_s l_s^{3}}\int_{{\cal B}_2}\sqrt{ \det \left ( g -\frac{F_2}{F_0} \right ) }\ .
\end{equation}
where we used the tadpole cancellation condition.  We will take the cycle ${\cal B}_2$ to be a representative of the non--trivial cycle, which is the diagonal of the two $S^2$'s. Using the explicit form for the metric in (\ref{metric6def}),  as well as (\ref{eq:fluxes}), (\ref{eq:c}) and  (\ref{eq:coeffs}), we get:
\begin{equation}\label{premass}
m_{D2} L = 4 \pi  \frac{n_0 L}{(2\pi)^2 g_s l_s^{3}}  \sqrt{ \left ( \frac{e^{2B_1}}{4} \right )^2 +\left ( \frac{k_2 e^{2B_1}}{F_0} \right )^2 } = 2 \pi^2  \frac{n_0 l^3}{g_s}  w_0(\psi_1) \sqrt{\frac{1+\psi_1^2}{\psi_1^2}}\ .
\end{equation}
The fact that the two expressions under the square root are proportional is related to the BPS condition, as
discussed in appendix \ref{sub:bpsours}.

Inserting the values of $l$ and $g_s$ from (\ref{parameters}),(\ref{parameterstwo}) for generic fluxes we obtain
\begin{equation}
m_{D2} L = \left ( \frac{n_2^2}{2} - n_0 n_4 \right ) \frac{2 \pi^2 w_0(\psi_1)}{ f_4(\psi_1)} \sqrt{\frac{1+\psi_1^2}{\psi_1^2}}\,.  \end{equation}
Quite remarkably, the function of $\psi_1$ in the previous formula, which can be computed numerically,  turns out to be  constant with value $1$. The final result for the mass formula is then
 \begin{equation}\label{mass}
m_{D2} L =   \left ( \frac{n_2^2}{2} - n_0 n_4 \right )\ .
 \end{equation}
Upon using the dictionary (\ref{eq:dic}),
this formula is identical to the field theory prediction (\ref{eq:Danielmonopoles}) in the limit where $n_0\ll n_2$. This is exactly the limit where we can trust the supergravity solution, since, as shown in (\ref{parameters}), for a generic value of $\psi_1$, $L$ is large only if $n_2\gg n_0$. In this limit, it is also true that the dimension of the corresponding operator is given by $\Delta \sim m_{D2} L$.

In contrast with the ${\cal N}=1$ results in section \ref{sub:probes1}, and with the naive expectation expressed earlier, we see from (\ref{mass}) that the mass of the bound state remains finite also  in the limit $N\gg k^3/n_0^2$. A contribution from the $B$ field cancels the large mass $\sim L/g_s$ of the constituent D0-brane, leaving a smaller piece that is proportional to the volume of the shrinking $S^2$. These are precisely the new light states that we had predicted to exist from the field theory analysis in section \ref{sub:monoABJM}.

% subsection probesII (end)

% section n2 (end)

\section*{Acknowledgements}

O.~A.~is supported in part by the Israel--U.S.~Binational Science Foundation, by a  research center supported by the Israel Science Foundation (grant number 1468/06), by a grant (DIP H52) of the German Israel Project Cooperation, and by the Minerva foundation with funding from the Federal German Ministry for Education and Research. D.~J.~wishes to acknowledge funding provided by the Association of Members of the Institute for Advanced Study. A.~T.~and  A.~Z.~are supported in part by INFN and MIUR under contract 2007--5ATT78--002.

\begin{appendix}

\section{Supersymmetry equations and pure spinors for the ${\cal N}=2$ solution}% (fold)
\label{apppure}

We will give in this section more details about the ${\cal N}=2$ solution we found in section \ref{sec:n2}.

The supersymmetry parameters for compactifications of the form ${\rm AdS}_4 \times M_6$ (or ${\rm Minkowski}_4 \times M_6$) decompose as
\beq
\label{spinan}
\begin{split}
    \epsilon_1 =\sum_{a=1}^{{\cal N}}\zeta^a_+ \otimes \eta^{1a}_+ +  \zeta^a_- \otimes \eta^{1a}_- \ ,\\
    \epsilon_2 = \sum_{a=1}^{{\cal N}} \zeta^a_- \otimes \eta^{2a}_+ +  \zeta^a_+ \otimes \eta^{2a}_-  \ .
\end{split}
\eeq
Here, ${\cal N}$ is the number of supersymmetries. The subscripts $\pm$ denote
positive and negative chirality spinors, in four and six dimensions; the negative chirality spinors are conjugate to the positive chirality ones,
\begin{equation}\label{eq:spincon}
    \zeta^a_- = (\zeta^a_+)^* \ ,\qquad
    \eta^{ia}_- = (\eta^{ia}_+)^*\ .
\end{equation}
For each $a$, $\zeta^a_+$ can vary among a basis of four--dimensional Weyl spinors; we will take the elements of this basis to be ``Killing spinors'', which means that $D_\mu \zeta_+ = \frac\mu2 \gamma_\mu \zeta_-$.  The $\eta^{ia}_+$, with $i=1,2$,
are a priori independent six--dimensional Weyl spinors.
In this section, we will consider ${\cal N}=2$.

A priori, one could have taken the $\zeta^a$ in $\epsilon_1$ and $\epsilon_2$ to be different. This can indeed be done for compactifications with vanishing RR flux; for example, for the usual ${\cal N}=2$ Calabi--Yau compactifications. To recover that case in (\ref{spinan}), one can take for example $\eta^{21}=\eta^{12}=0$, and keep a non--vanishing $\eta^{11}$ and $\eta^{22}$. However, in compactifications where RR fluxes are present, the $\zeta^a$ in $\epsilon_1$ and $\epsilon_2$ are required to be equal, up to a constant that can be reabsorbed in the $\eta^{ia}$. Hence (\ref{spinan}) describes all possible ${\cal N}=2$ compactifications, and is particularly appropriate for vacua with RR fluxes.

Using (\ref{spinan}) in the supersymmetry equations yields equations for the internal spinors $\eta^{ia}$. In fact, these equations do not mix the $\eta^{i1}$ with the $\eta^{i2}$. In what follows, we will first analyze the equations of the $\eta^{i1}\equiv \eta^i$; we will come back to the second pair later.

We will construct a pair of pure spinors as tensor products of the supersymmetry parameters\footnote{\label{foot:cl}As usual, we left implicit a Clifford map on the left hand side, that sends $dx^m\to \gamma^m$.} $\eta^1$ and $\eta^2$
\beq
\label{purespindef}
\Phi_\pm = \eta^1_+ \otimes \eta^{2 \, \dagger}_{\pm} \ .
\eeq

The type IIA supersymmetry conditions can be expressed as \cite{gmpt3}:
\begin{subequations}\label{eq:susy}
    \begin{align}
        \label{susyeq1app}
        & ({\rm d}-H\wedge) (e^{A-\varphi} \re(\Phi_-) ) = 0 \ ,\\
        \label{susyeq2app}
        & ({\rm d}-H\wedge) (e^{3A-\varphi} \im (\Phi_-) ) = - 3 e^{2A - \varphi} \mu \im(\Phi_+)  + \frac{e^{4 A}}{8} \ast \lambda(F) \ ,\\
        \label{susyeq3app}
        & ({\rm d}-H\wedge) (e^{2A-\varphi}  \Phi_+)= - 2  \mu  e^{A - \varphi}  \re(\Phi_-)  \ ; \\
        \label{eq:susynorm}
        &|| \Phi_+ || = || \Phi_- || = e^A\ .
    \end{align}
\end{subequations}
Here, $F$ are the internal fluxes (which determine also the external fluxes, by self--duality). $A$ is the warping function, defined as $ds^2_{10}=e^{2A}ds^2_{{\rm AdS}_4} + ds^2_6$. The cosmological constant in $ds^2_{{\rm AdS}_4}$ is given by $\Lambda =- 3 |\mu|^2$. Since $A$ is non--constant in the solution, however, this $\Lambda$ has no independent meaning, since one can reabsorb it in $A$. We have normalized $\mu=2$ in this paper. The symbol $\lambda$ acts on a $k$--form by multiplying it by the sign $(-)^{{\rm Int}(k/2)}$.
Finally, the norm in (\ref{eq:susynorm}) is defined as $|| A||^2 = i (A\wedge \lambda(\bar A))_6$.

The metric (\ref{metric6def}) can be written in terms of the vielbein
\begin{equation}\label{eq:Ei}
    \begin{split}
        E_1 &=  \frac{1}{2\sqrt{2}} \epsilon dt+  \frac{i}{8} \Gamma (da +A_2-A_1), \\
        E_2 &= \frac12 \left ( e^{B_1} \sin (t)    e^{-i a/2}e_1 - e^{B_2} \cos (t) e^{i a/2}  e_2 \right ), \\
        E_3 &= \frac12 \left (e^{B_1} \cos (t)  e^{-i a/2}  e_1+ e^{B_2} \sin (t)e^{i a/2}   e_2 \right ),
    \end{split}
\end{equation}
where $e_i=d\theta_i+i \sin\theta_i d\phi_i$ are the natural one--forms on the spheres $S^2_i$. For $e^{B_1}=\cos(t)$ and
$e^{B_2}=\sin(t)$ we recover the Fubini--Study metric of $\mathbb{CP}^3$ with natural K\"ahler form  $J= \frac{i}{2} \sum_{i=1}^3 E_i\wedge \bar E_i $  and natural three form section $\Omega = E_1\wedge E_2\wedge E_3$ (see for example \cite[(5.31)]{gaiotto-t2}). It is also convenient to use the forms
\begin{equation}\label{eq:o}
    J_i=  d A_i=\frac{i}{2} e_i\wedge \bar e_i \ \ {\rm (not\  summed)} \ ,\qquad
    o \equiv \frac{i}{2} e^{i a} e_2\wedge \bar e_1\ ;
\end{equation}
the $J_i$ were already defined in (\ref{eq:dAJ}). These forms satisfy
\begin{equation}\label{eq:do}
    d J_i=0 \ ,\qquad d o = i (da+A_2-A_1)\wedge o \ ,\qquad
    o \wedge \bar o = -J_1 \wedge J_2\ .
\end{equation}

The generic pure spinors corresponding to an ${\rm SU}(3) \times {\rm SU}(3)$ structure can be written in terms of the ``dielectric Ansatz'' \begin{equation}\label{susyI}
    \begin{split}
        \Phi_+ &= \frac{i}{8} \cos (2\psi) \, e^{A+i\theta} \, {\rm exp}\left (- i\left ( \frac{j}{\cos (2\psi)} + \frac{i}{2} z\wedge \bar z\right ) + \tan (2 \psi) \, {\rm Re} (\omega)\right )\ ,\\
        \Phi_- &= - \frac{i}{8} \sin (2 \psi)e^{A+i\theta} \,  z  \wedge  \, {\rm exp}\left (- \cot (2\psi) \, {\rm Re} (\omega) - \frac{i}{\sin (2\psi)} {\rm Im} (\omega) \right )\ ,
    \end{split}
\end{equation}
where $\theta$ and $\psi$ are two new angular variables; one can see easily that the supersymmetry equations (\ref{eq:susy}) relate them by
\begin{equation}\label{eq:theta}
    \tan (\theta) =- \cot (2 t) \, \sin (2\psi)\ .
\end{equation}
The one--form $z$ and the two--forms $j$ and $\omega$ can also be used to describe an SU(2) structure on $M_6$. For our solution, these forms are given by
\begin{equation}\label{eq:su2}
    \begin{split}
        z &= - i e^{-i\theta}E_1\ ,\\
        j & = \frac{i}{2} \left (E_2 \wedge \bar E_2 + \bar E_3\wedge E_3 \right )\ , \\
        \omega &= i E_2 \wedge \bar E_3\ .
    \end{split}
\end{equation}
We can also characterize $j$ and $\omega$ in terms of the forms in (\ref{eq:o}):
\begin{equation}\label{eq:jot}
    \left(\begin{array}{cc}\vspace{.3cm}
        j \\ - {\rm Re} (\omega)
    \end{array}\right) =
    \frac14
    \left(\begin{array}{cc}\vspace{.3cm}
        \cos(2t) & - \sin(2t)\\
        \sin(2t) & \cos(2t)
    \end{array}\right)
    \left(\begin{array}{cc}\vspace{.3cm}
        -e^{2B_1} J_1 + e^{2B_2} J_2 \\ 2 e^{B_1 + B_2} {\rm Re} (o)
    \end{array}\right)
     \ ,\quad
    {\rm Im} (\omega) = \frac 12 e^{B_1 + B_2} {\rm Im}  (o) \ .
\end{equation}

The RR fluxes are determined to be as in equation (\ref{eq:fluxes}) with
\begin{equation}\label{eq:coeffs}
    \begin{split}
        k_2 &= \frac{ c \, e^{-4A}}{ 2 \,w_1 } \frac{\sec (2\psi)}{\cos(2t)} \left ( 2\, C_{t,\psi} +  w_1 \right ),\\
        g_2 &= - \frac{ c\,  e^{-4A}}{ 2 \,w_2 } \frac{\sec (2\psi)}{\cos(2t)}\left ( 2\, C_{t,\psi} +  w_2 \right ),\\
        \tilde k_2 &= 2\,\frac{ c \, e^{-4A}}{ w_1\, w_2} \left ( 2\,C_{t,\psi}(w_1+w_2) + 3\,  w_1\,  w_2 \right ),\\
        k_4 &= -\frac{ c \, e^{-4A}}{4 \, w_1\,  w_2} \frac{ \sin (2\psi) }{\sin(4t) \cos^2(2 \psi)} \left ( 2\,C_{t,\psi}(w_1+w_2) + w_1 \, w_2 \right ),\\
        \tilde k_4 &= \frac{ c \, e^{-4A}}{2 \, w_2 } \frac{\tan (2\psi)}{\sin(2t)} \left ( 2 \,C_{t,\psi} + 3 \, w_2 \right ),\\
        \tilde g_4 &= -\frac{ c \, e^{-4A}}{2 \, w_1 }\frac{\tan (2\psi)}{\sin(2t)} \left ( 2 \,C_{t,\psi} + 3 \, w_1 \right ),\\
        k_6 &= 6\, c \, e^{-4 A},
    \end{split}
\end{equation}
where $C_{t,\psi}$ was defined in (\ref{eq:Ctpsi}).
Recall also that one possible choice of NS--NS field that satisfies the equations of motion is $B_0 = F_2/F_0$, as in (\ref{eq:f2t0}).

So far we have described the solution as if it were an ${\cal N}=1$ solution: we have only paid attention to the $a=1$ part of (\ref{spinan}). To show that the solution actually has ${\cal N}=2$ supersymmetry, we have to provide a second pair of spinors, $\eta^{i2}$, that satisfies the equations of motion for supersymmetry with the same expectation values for all the fields. In terms of pure spinors, we can now form the bilinears
\begin{equation}
    \tilde\Phi_\pm = \eta^{12}_+ \otimes \eta^{22\,\dagger}_\pm\
\end{equation}
and require that they solve again the equations (\ref{eq:susy}), with the same values of the fluxes and the same metric.

In fact, one expects the two solutions $\Phi$ and $\tilde \Phi$ to be rotated by R--symmetry, so that there is actually a ${\rm U}(1)$'s worth of solutions to (\ref{eq:susy}). To see this ${\rm U}(1)$, rotate the two--form $o$ in (\ref{eq:o}) by a phase:\footnote{Alternatively, one can change the vielbeine (\ref{eq:Ei}) by translating $a \to a + \alpha$.}
\begin{equation}\label{eq:oa}
    o \to e^{-i \alpha} o \equiv o_\alpha\ .
\end{equation}
We can correspondingly define a pair of pure spinors $\Phi_{\pm}^\alpha$, by changing $o\to o_\alpha$ wherever it appears. The crucial fact about the rotation of $o$ in (\ref{eq:oa}) is that it keeps its differential properties (\ref{eq:do}) unchanged: namely, $d o_\alpha = i (d a + A_2 - A_1) \wedge o_\alpha$.  Because of this fact, the computations to check (\ref{eq:susy}) do not depend on $\alpha$; and, since we checked already that $\alpha=0$ gives a solution, it follows that any $\Phi_\pm^\alpha$ is a solution. A priori, this could be a solution with different fluxes; but we can see from (\ref{eq:fluxes}) that $o$ never appears in $F_k$. We conclude, then, that the solution we have found is an ${\cal N}=2$ solution.
% section apppure (end)

\section{BPS particles} % (fold)
\label{sec:bps}

In this section, we will give a general analysis of BPS particles in flux compactifications (subsection \ref{sub:bpsgen}), and we will then apply those general results to the ${\cal N}=2$ background described in section \ref{sec:n2} and appendix \ref{apppure}.

\subsection{General considerations} % (fold)
\label{sub:bpsgen}

We will start with some general considerations about BPS states in ${\cal N}=2$ backgrounds with fluxes. These will in general be states that are left invariant by a certain subalgebra of the supersymmetry algebra. This subalgebra is in general defined by the fact that the two supersymmetry parameters $\epsilon_i$ are related:
\begin{equation}\label{eq:par}
    \Gamma_\parallel \epsilon_2 = \epsilon_1 \ .
\end{equation}
In first approximation, $\Gamma_\parallel$ is the product of the gamma matrices parallel to the brane. When $B$ fields or worldsheet fluxes ${\cal F}$ are present, $\Gamma_\parallel$ receives additional contributions of $e^{{\cal F}-B}$. We will give a definition later on, in the context needed for this paper; for the general and explicit expression, see for example \cite[Eq.~(3.3)]{martucci-smyth}.
For an ${\rm AdS}_4 \times M_6$ compactification, we would like to use the decomposition (\ref{spinan}). For particles, this will lead to an equation involving the four--dimensional spinors $\zeta^i_\pm$ and $\gamma_0 \zeta^a_\pm$; here and in what follows, the index ${}_0$ is meant to be a frame index.  To have a chance to solve the resulting equations, we need to postulate a relation between these spinors. One can write for example
\begin{equation}\label{eq:g0ans}
    \gamma_0 \zeta^a_+ = A^{ab} \zeta^b_- \ ,
\end{equation}
for some matrix $A$. (Recall that in general the index $a$ runs from 1 to ${\cal N}$; for us, ${\cal N}=2$, and so $a=1,2$.) In fact, (\ref{eq:g0ans}) is almost the most general choice one can make, compatibly with the symmetries of the problem. The only generalization one could make would be to multiply the left-hand side by another matrix $B^{ab}$. Whenever this matrix is invertible, one can reabsorb it by a redefinition of $A^{ab}$. In this sense, we can say that (\ref{eq:g0ans}) is the ``generic'' Ansatz for a BPS particle.

The matrix $A^{ab}$ in (\ref{eq:g0ans}) needs to satisfy certain conditions. Let us work for simplicity in a basis where all the space--time gamma matrices $\gamma_\mu$, $\mu=0,\ldots,3$ are real, and the internal $\gamma^m$, $m=1,\ldots,6$, are purely imaginary; the ten--dimensional gamma matrices are then given as usual by
\begin{equation}\label{eq:gammadec}
    \Gamma_\mu = e^A \gamma_\mu \otimes 1 \ ,\qquad
    \Gamma_{m+3} = \gamma_5 \otimes \gamma_m \ .
\end{equation}
It follows from these definitions that $\Gamma_\mu$ are real. Let us now conjugate (\ref{eq:g0ans}); using (\ref{eq:spincon}), the fact that $\gamma_0$ is real, and that $\gamma_0^2 = -1$, we get
\begin{equation}\label{eq:AA}
    A^{ab}\overline{A^{bc}}= - \delta^{ac} \ .
\end{equation}
If we were considering an ${\cal N}=1$ background, $A^{ab}$ would be a one--by--one matrix, and (\ref{eq:AA}) would have no solution. This is just what one would expect: there are no BPS particles in a ${\cal N}=1$ background. For ${\cal N}=2$, one choice that satisfies (\ref{eq:AA}) is
\begin{equation}\label{eq:A}
    A= e^{-i \lambda} \left(\begin{array}{cc}\vspace{.3cm}
        0 & \ 1 \\ - 1 & \ 0
    \end{array}\right)\ .
\end{equation}

We can now use (\ref{eq:gammadec}) to write
\begin{equation}\label{eq:gammapardec}
    \Gamma_\parallel = \gamma_0 \otimes \gamma_\parallel\ ,
\end{equation}
where $\gamma_\parallel$ is now an element of the internal Clifford algebra; it contains the product of all the internal gamma matrices parallel to the brane, plus additional contributions from the worldsheet flux and $B$-field. Let ${\cal B}_p\subset M_6$ be the $p$--cycle wrapped by the brane, of dimension $p$ and with coordinates $\sigma^\alpha$, $\alpha=1,\ldots,p$. Then we define the natural volume form on ${\cal B}$ to be
\begin{equation}\label{eq:vol}
    {\rm vol}_{{\cal B}}\equiv \sqrt{\det(g+ {\cal F}-B)}\, d \sigma^1\wedge\ldots\wedge d \sigma^p\ .
\end{equation}
 One can also define similarly an ``inverse volume form'' as the multivector
\begin{equation}
    {\rm vol}_{\cal B}^{-1}= \frac{\del_1\wedge \ldots \wedge \del_p}{\sqrt{\det(g+ {\cal F}-B)}}\ ,
\end{equation}
which is a section of $\Lambda^p (T {\cal B})$. This multivector can be used to give an intrinsic definition of $\gamma_\parallel$: here is how. We can define $e^{{\cal F}}{\rm vol}^{-1}$ to be the multivector of mixed degree that one obtains by contracting the indices of the form $e^{\cal F}$ with the multi--vector ${\rm vol}^{-1}$. Recall now that multivectors can be ``pushed forward'': if we call  $x: {\cal B} \hookrightarrow  M_6$ the embedding map, with components $x^m(\sigma)$, then $x_*(e^{\cal F} {\rm vol}^{-1})$ is a multivector in $M_6$, obtained by contracting all indices $\alpha$ on ${\cal B}$ with the tensor $\del_\alpha x^m$. In fact:
\begin{equation}\label{eq:gammap}
    \gamma_\parallel = x_* (e^{{\cal F}-B} {\rm vol}_{\cal B}^{-1})\ .
\end{equation}
Here, we left implicit on the right hand side a Clifford map that sends a vector $\del_m$ into a gamma matrix $\gamma_m$. We already used this map on forms (see footnote \ref{foot:cl}). One can show that $\gamma_\parallel$ is unitary:
\begin{equation}\label{eq:gammaparunit}
    \gamma_\parallel^\dagger \gamma_\parallel = 1\ .
\end{equation}
For a more explicit expression of $\gamma_\parallel$,
see \cite[Eq.~(3.5)]{martucci-smyth}.

If we now use (\ref{eq:g0ans}), (\ref{eq:gammapardec}) and (\ref{spinan}) in (\ref{eq:par}), we get
\begin{equation}
    \gamma_\parallel\, \eta^{2a}_+ = (A^{-1})^{ba} \eta^{1b}_+ \ .
\end{equation}
For our choice (\ref{eq:A}), this reads
\begin{subequations}\label{eq:pareta}
    \begin{align}
        \label{eq:pareta22}
        \gamma_\parallel \, \eta^{22}_+ &= - e^{i \lambda} \eta^{11}_+\ ,\\
        \label{eq:pareta21}
        \gamma_\parallel \, \eta^{21}_+ &= e^{i \lambda} \eta^{12}_+\ .
    \end{align}
\end{subequations}

We are now left with solving (\ref{eq:pareta}), which are two purely internal equations. Each of the two equations is formally identical to others that have already appeared \cite{martucci-smyth} in the context of BPS objects which do exist in ${\cal N}=1$ flux compactifications: branes which extend along the time direction, plus one, two or three space directions. Hence we can simply follow the same steps; we will now summarize that procedure for (\ref{eq:pareta22}), and then apply the result to (\ref{eq:pareta21}).

Let us first define the new pure spinors
\begin{equation}\label{eq:Psi}
    \Psi_+ \equiv \eta^{11}_+ \otimes \eta^{22\,\dagger}_\pm\ ;
\end{equation}
notice that these are different from the pure spinors $\Phi_\pm$, defined in (\ref{purespindef}), which entered the supersymmetry equations (\ref{eq:susy}). In (\ref{purespindef}), $\eta^1$ and $\eta^2$ were to be understood as $\eta^{1a}$ and $\eta^{2a}$, for $a$ either 1 or 2. In (\ref{eq:Psi}), we are mixing $a=1$ with $a=2$.

A possible basis for the space of spinors of positive chirality is given by $\eta^{11}_+$ and $\gamma^m \eta^{11}_-$. Three linear combinations of the $\gamma^m$ make $\gamma^m \eta^{11}_-$ vanish: they are its three ``annihilators'' $\gamma^i$, where $i$ is a holomorphic index with respect to an almost complex structure $I$. Explicitly we have $\eta^{11\,\dagger}_+ \gamma_m \gamma^n \eta^{11}_+= (1+ iI)_m{}^n\equiv 2 \bar \Pi_m{}^n$. In terms of this basis a priori one can expand
\begin{equation}\label{eq:ab}
    \gamma_\parallel \eta^{22}_+ = a \eta^{11}_+ + b_m \gamma^m \eta^{11}_- \ .
\end{equation}
The coefficients $a$ and $b_m$ have a geometrical interpretation. To compute $a$, we can multiply (\ref{eq:ab}) from the left by $\eta^{11\,\dagger}_+$; we get $a e^A= \eta^{11\,\dagger}_+ \gamma_\parallel \eta^{22}_+= {\rm Tr}(\gamma_\parallel \eta^{22}_+ \eta^{11\,\dagger}_+)= {\rm Tr}(\gamma_\parallel \Psi_+^\dagger)$. From the formula
\begin{equation}
    {\rm Tr}(\sla A \sla B^\dagger) = \frac 8{k!} A_{m_1\ldots m_k}\bar B^{m_1\ldots m_k}  \ ,
\end{equation}
we see that ${\rm Tr}(\gamma_\parallel \Psi_+^\dagger)$ consists of contracting the free indices of $\gamma_\parallel$ with those of $\bar\Psi_+$. From (\ref{eq:gammap}), we see that $\gamma_\parallel$ contains factors of $\del_\alpha x^m$; when contracting with $\bar \Psi_+$, these factors reconstruct a pull--back of that form. In conclusion we get\footnote{The factor of $e^A$ comes from the fact that $\forall a, i$, $|| \eta^{ia} ||= e^{A/2}$, which follows from (\ref{eq:susy}).\label{foot:eA}}
\begin{equation}\label{eq:a}
    (e^{{\cal F}-B}\bar\Psi_+)|_{\cal B} = \frac a8 e^A\, {\rm vol}_{\cal B}\ ,
\end{equation}
where $|_{\cal B}$ denotes the top--form part on ${\cal B}$ of the pull--back. By similarly multiplying (\ref{eq:ab}) from the left by $\eta^{11\,\dagger}_- \gamma_n$, we get
\begin{equation}\label{eq:bi}
    (dx^m\cdot e^{{\cal F}-B} \Psi_-)|_B = -\frac14 b^n \Pi_n{}^m e^A\, {\rm vol}_{B}\ ,
\end{equation}
where $\cdot$ denotes the Clifford product: $v\cdot= v\wedge + v\llcorner$. Here, $\del_m \llcorner (dx^{m_1}\wedge \ldots \wedge dx^{m_p})\equiv p\,\delta_m{}^{[m_1}dx^{m_2}\wedge \ldots dx^{m_p]}$.

We can now go back to (\ref{eq:pareta22}). Comparing to the expansion (\ref{eq:ab}), we get
\begin{equation}
    a=-e^{i\lambda}\ ,\qquad b_m=0\ .
\end{equation}
Using the geometrical interpretations (\ref{eq:a}) and (\ref{eq:bi}), we get
\begin{equation}\label{eq:Psivol}
    {\rm Re} (-e^{-i \lambda } e^{{\cal F}-B}\Psi_+)|_B=\frac18 e^A {\rm vol}_B\ ,
\end{equation}
and
\begin{subequations}\label{eq:Psi0}
    \begin{align}\label{eq:Psi+0}
        &{\rm Im} (e^{-i \lambda } e^{{\cal F}-B}\Psi_+)|_B=0, \\
        & (v \cdot e^{{\cal F}-B}\Psi_-)|_B = 0 \ . \label{eq:Psi-0}
    \end{align}
\end{subequations}

Actually, one can show that (\ref{eq:Psivol}) is equivalent to the system (\ref{eq:Psi0}). To see this, observe that $\gamma_\parallel$ is unitary, as we saw in (\ref{eq:gammaparunit}). This implies that $\gamma_\parallel \eta^{22}_+$ should have the same norm as $\eta^{22}_+$.  Since all the spinors have norm $e^A$ (see footnote \ref{foot:eA}), it follows that
\begin{equation}
    |a|^2 + 2 b_m \bar b^m=1\ .
\end{equation}
This means that imposing ${\rm Re}(a)=1$ is equivalent to imposing ${\rm Im}(a)=0$ and $b_m=0$. Recalling (\ref{eq:a}) and (\ref{eq:bi}), we get our claim that (\ref{eq:Psivol}) is equivalent to (\ref{eq:Psi0}).

This completes our analysis of (\ref{eq:pareta22}) (along the lines of \cite{martucci-smyth}). For (\ref{eq:pareta21}), similar considerations apply; we obtain
\begin{equation}\label{eq:tPsivol}
    {\rm Re} (e^{-i \lambda } e^{{\cal F}-B}\tilde\Psi_+)|_B= \frac 18 e^A {\rm vol}_B\ ,
\end{equation}
and
\begin{subequations}\label{eq:tPsi0}
    \begin{align}\label{eq:tPsi+0}
        &{\rm Im} (e^{-i \lambda } e^{{\cal F}-B}\tilde\Psi_+)|_B=0, \\
        & (v \cdot e^{{\cal F}-B}\tilde\Psi_-)|_B = 0 \ , \label{eq:tPsi-0}
    \end{align}
\end{subequations}
for the pure spinors
\begin{equation}\label{eq:Psit}
    \tilde \Psi_\pm \equiv \eta^{12}_+ \otimes \eta^{21\,\dagger}_\pm\ .
\end{equation}

Let us now summarize this section: we have shown that a brane wrapping an internal cycle ${\cal B}$, and extended along the time direction, is BPS if and only if (\ref{eq:Psivol}) (or equivalently (\ref{eq:Psi0})) is satisfied by $\Psi$ and, analogously, (\ref{eq:tPsivol}) (or equivalently (\ref{eq:tPsi0})) is satisfied by  $\tilde \Psi$, where $\Psi$ and $\tilde\Psi$ are defined respectively in
(\ref{eq:Psi}) and (\ref{eq:Psit}). We will now compute these pure spinors for the solution described in section \ref{sec:n2} and in appendix \ref{apppure}.

% subsection bpsgen (end)

\subsection{D2/D0 bound states in the ${\cal N}=2$ solution} % (fold)
\label{sub:bpsours}

%We now analyze the BPS properties of particles obtained by wrapping D2-branes in the ${\cal N}=2$ solution discussed in section \ref{sec:n2}.

As discussed in the previous section, in order to study the supersymmetry  of BPS particles obtained from wrapped branes,  we need to form bilinears in the supersymmetry spinors  $\eta_+^{ia}$. We first need to write them explicitly. A convenient basis to expand our spinors is given by the pair of spinors defining
the ${\rm SU}(3)\times {\rm SU}(3)$  structure in (\ref{eq:su2}). Recall that a $SU(3)$ structure is specified by two invariant tensors $(J,\Omega)$ or, equivalently, by a spinor $\eta_+$ (of norm 1) such that
\begin{equation}
    \begin{split}
        \eta_+ \otimes \eta_+^\dagger &= \frac{1}{8} e^{-i J}\ ,\\
        \eta_+ \otimes \eta_-^\dagger &=  - \frac{i}{8} \Omega\ .
    \end{split}
\end{equation}
The ${\rm SU}(3)\times {\rm SU}(3)$  structure in (\ref{eq:su2}) can be seen as  the intersection of two $SU(3)$ structures given by $(J_1,\Omega_1) =(j+\frac{i}{2} z\wedge \bar z,  \omega\wedge z)$ and  $(J_2,\Omega_2) =(-j+\frac{i}{2} z\wedge \bar z, -\bar \omega\wedge z)$. We call the corresponding spinors $\eta_+$ and $\chi_+$. They are related by $ \chi_+ = \frac{1}{\sqrt{2}} z \cdot \eta_-  $, where  $z \cdot$ denotes  the Clifford multiplication by the one-form $z_m \gamma^m$.  We will need in the following an expression for the tensor products of a generic linear combination
\begin{equation}
    \begin{split}
        \mu_+ &= a \, \eta_+ + b \, \chi_+\ ,\\
        \nu_+ &= x \, \eta_+ + y\,  \chi_+ \ .
    \end{split}
\end{equation}
This is given by \cite{minasian-petrini-zaffaroni}
\begin{equation}
    \begin{split}
        \label{purespSU2}
        & \mu_+ \otimes \nu_+^\dagger = \frac{1}{8} \Big[ a \bar x e^{- i j} +  b \bar y e^{i j} - i
        ( a \bar{y} \omega + \bar{x} b \bar{\omega} ) \Big] e^{1/2 z \bar{z}} \ ,  \\
        & \mu_+ \otimes \nu_-^\dagger   = \frac{1}{8} \Big[ i (b y \bar{\omega} - a x \omega) + ( b x
        e^{i j} - a y e^{- i j}) \Big] z\ .
    \end{split}
\end{equation}
We can choose the spinors for the first supersymmetry as follows
\begin{equation}
    \begin{split}
        \eta^{11}_+ &= i e^{A/2+i\theta} ( e^{i \frac{\pi}{4}} \cos(\psi) \, \eta_+ - i e^{-i \frac{\pi}{4}} \sin(\psi)\,  \chi_+)\ , \\
        \eta^{21}_+ &=  i e^{A/2} (e^{i \frac{\pi}{4}} \cos(\psi) \, \eta_+ +  i e^{-i \frac{\pi}{4}} \sin(\psi) \, \chi_+)\ .
    \end{split}
\end{equation}
It is  easy to reproduce,  using formula (\ref{purespSU2}),  the dielectric ansatz (\ref{susyI}) for the pure spinors .

As discussed in appendix \ref{apppure}, there is a $U(1)$ family of supersymmetries obtained by rotating $o\rightarrow o_\alpha=e^{-i\alpha} o$. We can conveniently choose as a second independent supersymmetry  the one with
$o_{\pi}= -o$. This is defined by
\begin{equation}
    \begin{split}
        \eta^{12}_+ &=  i e^{A/2+i\theta} (e^{i \frac{\pi}{4}} \cos(\psi) \, \tilde \eta_+ - i e^{-i \frac{\pi}{4}} \sin(\psi)\,  \tilde \chi_+)\ , \\
        \eta^{22}_+ &=  i e^{A/2} (e^{i \frac{\pi}{4}} \cos(\psi) \, \tilde \eta_+ +  i e^{-i \frac{\pi}{4}} \sin(\psi) \, \tilde\chi_+ )\ ,
    \end{split}
\end{equation}
where
\begin{equation}
    \begin{split}
        \tilde\eta_+ &=   - i \cos (2 t)\,  \eta_+ + i \sin (2 t) \, \chi_+\ ,\\
        \tilde\chi_+ &= i \sin (2 t) \, \eta_+ + i \cos (2 t) \, \chi _+ \ .
    \end{split}
\end{equation}
This reproduces  the rotated pure spinors $\Phi_\pm^\pi$.

With these ingredients, we can compute the spinors $\Psi_\pm$ and $\tilde \Psi_\pm$ defined in (\ref{eq:Psi}) and (\ref{eq:Psit}) and check the BPS conditions for a D2-brane. It is easy to see that the D2-brane considered in section \ref{sec:n2},  which wraps the diagonal $S^2$  and sits at the North or South pole,  is indeed supersymmetric. Let us consider, for definiteness, the North pole. At $t=0$, $\psi=0$ and we see that $\eta_+^{i2}= -i \eta_+^{i1}$. As a consequence, at $t=0$,
\begin{equation}
    \Psi_\pm=\pm i \Phi_\pm \ ,\qquad \tilde \Psi_\pm=- i \Phi_\pm\ ,
\end{equation}
 and we are reduced to check expressions for the pure spinors $\Phi_\pm$ at the North pole. Taking into account that $\psi=0$ there, we have
\begin{equation}
    \begin{split}
        &\Phi_+|_{t=0}= \frac i8 e^{A+i \theta} e^{-i J}\ , \\
        &\Phi_-|_{t=0}= \frac i8 e^{A+i\theta} z \wedge \omega\ .
    \end{split}
\end{equation}
The condition
(\ref{eq:Psi-0}) for $\Psi_-$ (and the analogous  (\ref{eq:tPsi-0})  for $\tilde\Psi_-$) gets contributions only from the contraction with the vector $z$ and it is automatically satisfied because $\omega$ vanishes at the North pole, $t=0$.
 It is easily seen that the conditions for $\Psi_+$ and $\tilde\Psi_+$ are equivalent  and it is enough to analyze those for $\Psi_+$.  Equation (\ref{eq:Psi+0})  reads
\beq
{\rm Im} \left (e^{i(\theta-\lambda)} e^{-i j}\right ) \wedge e^{{\cal F}-B}|_{{\cal B}_2} =  0,
\eeq
and determines the world-volume field
\beq
{\cal F} = (B+ \cot (\theta -\lambda) j)|_{{\cal B}_2}.
\eeq
We see that a wrapped D2 brane can be made supersymmetric by choosing an appropriate world-volume field.
However, as discussed in section \ref{sub:probes2}, to have a consistent BPS state we need to impose the
quantization of the world-volume field and the cancellation of tadpoles. As discussed there, the quantization condition
requires to take $n_0$ D2-branes. On the other hand, the tadpole condition requires ${\cal F}=\beta$ or, equivalently,  ${\cal F}-B= -F_2/F_0$. At $t=0$, using the explicit form for the metric in (\ref{metric6def}),  as well as (\ref{eq:fluxes}), (\ref{eq:c}), (\ref{eq:su2}), (\ref{eq:theta}) and (\ref{eq:coeffs}), we evaluate
$\tan(\theta) =-\psi_1$ and $j=-\frac14 e^{2B_1} J_1$ and $F_2/F_0=-\frac14 e^{2B_1} J_1/\psi_1$. Recall that $J_1$ is the volume form of one of the two $S^2$'s, as defined in (\ref{eq:dAJ}).  We thus see that the tadpole condition is satisfied by $\lambda=0$. The mass of $n_0$ D2 branes is then obtained by integrating the volume form in  (\ref{eq:Psivol})
\beq
n_0  \int_{{\cal B}_2} \sqrt{\det(g+ {\cal F}-B)}= n_0 \frac{1}{\sin(\theta)}   \int_{{\cal B}_2} J = n_0  \frac14 e^{2 B_1} \sqrt{\frac{1+\psi_1^2}{\psi_1}} \int_{{\cal B}_2} J_1\ .
 \eeq
Using this, one exactly reproduces the result (\ref{premass}) of section \ref{sub:probes2}.

A more detailed analysis of equations (\ref{eq:Psi0}) (and the analogous ones for $\tilde\Psi_\pm$) shows that
a D2-brane sitting at $t\ne 0,\pi/2$ cannot be supersymmetric and simultaneously satisfy the tadpole condition.

\end{appendix}

%\bibliography{at}
%\bibliographystyle{at}

\providecommand{\href}[2]{#2}

\end{document}